\documentclass[twocolumn,pra,showpacs,superscriptaddress]{revtex4-1}
\pdfoutput=1
\usepackage{amsmath, amssymb}
\usepackage{graphics}
\usepackage{pstricks}
\usepackage{graphicx,color}
\usepackage{hyperref}
\usepackage{verbatim}

\begin{document}

\title{Relaxation dynamics of the Lieb--Liniger gas following an interaction quench: \\ A coordinate Bethe-ansatz analysis}

\author{Jan C. Zill}
\email{j.zill@uq.edu.au}
\affiliation{School of Mathematics and Physics, The University of Queensland, Brisbane QLD 4072, Australia}
\affiliation{Institut f\"ur Theoretische Physik, Universit\"at Heidelberg, Philosophenweg 16, 69120 Heidelberg, Germany}
\affiliation{ExtreMe Matter Institute EMMI, GSI Helmholtzzentrum f\"ur Schwerionenforschung, 64291 Darmstadt, Germany}

\author{Tod M. Wright}
\affiliation{School of Mathematics and Physics, The University of Queensland, Brisbane QLD 4072, Australia}

\author{Kar\'en V. Kheruntsyan}
\affiliation{School of Mathematics and Physics, The University of Queensland, Brisbane QLD 4072, Australia}

\author{Thomas Gasenzer}
\affiliation{Institut f\"ur Theoretische Physik, Universit\"at Heidelberg, Philosophenweg 16, 69120 Heidelberg, Germany}
\affiliation{ExtreMe Matter Institute EMMI, GSI Helmholtzzentrum f\"ur Schwerionenforschung, 64291 Darmstadt, Germany}

\author{Matthew J. Davis}
\affiliation{School of Mathematics and Physics, The University of Queensland, Brisbane QLD 4072, Australia}

\begin{abstract}
We investigate the relaxation dynamics of the integrable Lieb--Liniger model of contact-interacting bosons in one dimension following a sudden quench of the collisional interaction strength.  The system is initially prepared in its noninteracting ground state and the interaction strength is then abruptly switched to a positive value, corresponding to repulsive interactions between the bosons.  We calculate equal-time correlation functions of the nonequilibrium Bose field for small systems of up to five particles via symbolic evaluation of coordinate Bethe-ansatz expressions for operator matrix elements between Lieb--Liniger eigenstates.  We characterize the relaxation of the system by comparing the time-evolving correlation functions following the quench to the equilibrium correlations predicted by the diagonal ensemble and relate the behavior of these correlations to that of the quantum fidelity between the many-body wave function and the initial state of the system.  Our results for the asymptotic scaling of local second-order correlations with increasing interaction strength agree with the predictions of recent generalized thermodynamic Bethe-ansatz calculations.  By contrast, third-order correlations obtained within our approach exhibit a markedly different power-law dependence on the interaction strength as the Tonks--Girardeau limit of infinitely strong interactions is approached. 
\end{abstract}

\pacs{02.30.Ik, 67.85.-d, 05.30.Jp}

\date{\today}

\maketitle

\section{Introduction}
Experiments in ultracold atomic physics offer the opportunity to study many-body quantum systems that are well isolated from their environment and exhibit dynamical evolution on observable time scales.  Moreover, the excellent control of trapping geometries now attainable in experiments allows for the near-direct realization of idealized models of condensed-matter systems~\cite{Bloch2008}.  In particular, experiments on degenerate Bose gases in quasi-one-dimensional trapping geometries approach the conditions assumed in the Lieb--Liniger (LL) model~\cite{Lieb1963a,*Lieb1963b} of indistinguishable bosons in one dimension (1D) interacting via a point interparticle potential~\cite{Olshanii1998,Petrov2000,Lieb2003,Cazalilla2011}.  The LL model plays an important role in the literature as a comparatively transparent, prototypical example of the class of quantum integrable models~\cite{Korepin1993,Sutherland2004}, which admit formal solutions in terms of the Bethe ansatz~\cite{Bethe1931}.   Experimental investigations of nonequilibrium dynamics with ultracold atoms have demonstrated the breakdown of conventional thermalization in quasi-1D Bose gases~\cite{Kinoshita2006,Hofferberth2007,Gring2012,Langen2013}.  These observations have fueled a rapidly growing program of theoretical research into the role of conservation laws in constraining the nonequilibrium dynamics of integrable systems in particular and the mechanisms of relaxation and origins of thermal equilibrium in isolated quantum systems in general~\cite{Cazalilla2010,Dziarmaga2010,Polkovnikov2011}.  
~\\

Theoretical works on the relaxation of integrable quantum systems initially focused on the class of spin chains and other interacting 1D systems that can be solved by a Jordan-Wigner transformation~\cite{Jordan1928} to a system of noninteracting fermions~\cite{Rigol2007,Pezer2007,Gangardt2008,Rossini2009,Rossini2010,Calabrese2011,Calabrese2012a,Calabrese2012b,Collura2013a,Collura2013b,Goldstein2013,Rigol2004,Rigol2006,Rigol2007,Cassidy2011,Gramsch2012,He2013,Wright2014}.   More recently, workers in this area have focused increasingly on the nonequilibrium dynamics and relaxation of the more general class of integrable quantum systems (such as the LL model) that can be solved by Bethe ansatz~\cite{Bethe1931} but do not admit a mapping to noninteracting degrees of freedom~\cite{Gasenzer2005,Berges2007,Branschadel2008,Gasenzer2008,Buljan2008,*Buljan2009,Jukic2008,Pezer2009,Fioretto2010,Kronenwett2011,Lamacraft2011,Iyer2012,Iyer2013,Mathy2012,Sato2012,Kaminishi2013,Mossel2010,Fagotti2013,Pozsgay2013}.  The quantum quench consisting of an abrupt change of the interparticle interaction strength of the LL model has recently emerged as an important test bed for theories of relaxation of such systems.  Such a scenario may be realized experimentally by making use of confinement-induced resonances~\cite{Olshanii1998,Bergeman2003,Haller2010}.  In this article we undertake calculations within the coordinate Bethe-ansatz formalism to investigate the dynamics following a quench of the interaction strength in small LL systems of at most five particles.

Results for the relaxation dynamics of the LL model following an interaction-strength quench have previously been obtained in the limiting cases of quenches to the noninteracting limit~\cite{Imambekov2009,Mossel2012a} and to the opposite Tonks--Girardeau (TG) limit of infinitely strong interactions~\cite{Gritsev2010,Kormos2014,Mazza2014,DeNardis2014b}, where the dynamics are governed by free-particle propagation.  For quenches to finite interaction strengths, the relaxation dynamics have been investigated using a range of techniques, including exact diagonalization within a truncated momentum-mode basis~\cite{Berman2004}, quasiexact numerical simulations of lattice discretizations of the model~\cite{Deuar2006,Muth2010a}, and nonperturbative approximations derived using functional-integral techniques~\cite{Gasenzer2005,Berges2007,Branschadel2008,Gasenzer2008,NoteA}.  A finite-size scaling analysis~\cite{Ikeda2013} of expectation values in energy eigenstates of the LL model indicated that the eigenstate thermalization hypothesis~\cite{Deutsch1991,Srednicki1994,Rigol2008} holds for this model in the weak sense~\cite{Biroli2010} only, implying the absence of thermalization following a quench.  A recently proposed generalization of the thermodynamic Bethe ansatz (TBA)~\cite{Mossel2012b,Caux2012} was used in Ref.~\cite{Kormos2013} to obtain the predictions of the nonthermal generalized Gibbs ensemble (GGE)~\cite{Jaynes1957a,*Jaynes1957b,Rigol2007} for the relaxed state following an interaction-strength quench.  This generalized TBA also forms the basis for the so-called quench-action variational approach~\cite{Caux2013,Mussardo2013}, which was used in Ref.~\cite{DeNardis2014a} to predict the dynamical evolution of correlation functions following such a quench.  We note also studies of related nonequilibrium scenarios such as a quench to the so-called super-Tonks regime~\cite{Astrakharchik2005,Haller2009} of strong attractive interactions~\cite{Chen2010,Muth2010b} and a coherent splitting~\cite{Gring2012} of the LL gas~\cite{Kaminishi2014}.  In higher dimensionalities, interaction quenches of Bose systems have been investigated within Bogoliubov-based~\cite{Carusotto2010,Rancon2013,Deuar2013,Natu2013,Sykes2014,Yin2013} theoretical descriptions, motivated in part by recent experiments on interaction-strength quenches in 2D~\cite{Hung2013} and 3D~\cite{Makotyn2014} Bose gases.  

In this article we undertake calculations within the coordinate Bethe-ansatz formalism to characterize the dynamics of the LL model following an interaction-strength quench.  Our methodology is based on the symbolic evaluation of overlaps and matrix elements between LL eigenstates in terms of the rapidities that label them.  The rapidities themselves are obtained by numerical solution of the appropriate Bethe equations.  Computational expense limits our calculations to small particle numbers $N\leq 5$.  However, our approach in terms of the exact eigenstates of the LL Hamiltonian explicitly respects the integrability of the model, in contrast to works that make use of lattice discretizations~\cite{Deuar2006,Muth2010a} of the LL Hamiltonian or explicit momentum-space cutoffs~\cite{Berman2004}.  Moreover, our approach allows us to calculate infinite-time averages of observables, i.e., expectation values in the so-called diagonal ensemble (DE)~\cite{Rigol2008}, in contrast to quasiexact numerical schemes that can only follow the relaxation dynamics for short time periods~\cite{Deuar2006,Muth2010a}.  

We observe clear signs of relaxation of the system to the DE in our results for dynamically evolving correlation functions, even for the small system sizes we consider. In particular, we calculate the time evolution of the momentum distribution of the Bose gas, which is not easily accessible within other Bethe-ansatz-based approaches~\cite{Kormos2014}, and find results qualitatively consistent with the results of functional-integral calculations of the relaxation dynamics~\cite{Gasenzer2005,Berges2007,Branschadel2008,Gasenzer2008,NoteA}.  Our results for the second-order coherence function reveal the propagation of correlation waves, as previously observed in simulations of quenches within lattice discretizations of the LL model~\cite{Deuar2006,Muth2010a} and quenches to the TG limit~\cite{Gritsev2010,Kormos2014}.   Our numerical approach in terms of the $N$-particle energy eigenstates of the LL Hamiltonian also allows us to calculate the quantum fidelity between the time-evolved state of the system following the quench and the initial state, which decays over time as the eigenstate dephasing that underlies the relaxation dynamics~\cite{Rigol2008} takes place.  We find, in particular, that the behavior of this fidelity is qualitatively similar to that of nonlocal quantities such as the occupation of the zero-momentum single-particle mode, indicating that these experimentally relevant quantities provide effective probes of the eigenstate dephasing of the $N$-body system.  

Our results for correlation functions in the DE are complementary both to exact analytic results for the stationary-state correlations following a quench to the TG limit~\cite{Kormos2014} and to the predictions of generalized thermodynamic ensembles for the equilibrium correlations following quenches to finite interaction strengths~\cite{Kormos2013,DeNardis2014a}.  For large interaction strengths, our results for the momentum distribution and static structure factor appear to be approaching the known TG-limit results~\cite{Kormos2014}.  Moreover, our results for second-order correlations in the DE corroborate the predictions of Refs.~\cite{Kormos2013,DeNardis2014a} for the generalized equilibrium state of the system.  In particular, our DE results for local second-order correlations are consistent with the power-law scaling with interaction strength predicted by Refs.~\cite{Kormos2013,DeNardis2014a}.  By contrast, however, we find that the power law with which local third-order correlations in the DE scale with interaction strength is markedly different from that predicted by these previous works, suggesting that further investigation of these correlations is necessary. 

This article is organized as follows.  Section~\ref{sec:model} contains a brief review of the LL model and the coordinate Bethe-ansatz approach to its solution, and outlines our methodology for the calculation of correlation functions within this formalism.  In Sec.~\ref{sec:dynamics} we present results on the time evolution of dynamical correlation functions following a quench of the interaction strength from the noninteracting limit to a finite repulsive value.  Section~\ref{sec:StatMech} compares the relaxed-state correlation functions, as described by the DE, to the predictions of conventional statistical mechanics and other theoretical approaches to the interaction-strength quench scenario.  In Sec.~\ref{sec:summary} we summarize our results and present our conclusions.

\section{Methodology}\label{sec:model}
\subsection{Lieb--Liniger model eigenstates}
The LL model~\cite{Lieb1963a,*Lieb1963b} describes a system of $N$ indistinguishable bosons subject to a delta-function pairwise interparticle interaction potential in a periodic 1D geometry.  In this article we work in units such that $\hbar=1$ and the particle mass $m=1/2$, and so the first-quantized Hamiltonian for this system can be written
\begin{equation}\label{eq:LLmodel}
    \hat{H} = - \sum_{i=1}^{N} \frac{\partial^2}{\partial x^2_i} + 2c \sum_{i<j}^{N} \delta(x_i - x_j),
\end{equation}
where $c$ is the interaction strength.  Hereafter, we restrict our attention to the case of non-negative interaction strengths $c\geq0$.  The solution of Hamiltonian~\eqref{eq:LLmodel} by Bethe ansatz was first described by Lieb and Liniger~\cite{Lieb1963a,*Lieb1963b}, and a detailed discussion of this approach can be found in Ref.~\cite{Korepin1993}.  For the reader's convenience, we provide a brief review of the method here.   

Due to the symmetry of the Bose wave function $\psi(\{x_i\})$ under the exchange of particle labels, it is (irrespective of the boundary conditions of the geometry) completely determined by its form on the fundamental permutation sector,
\begin{equation}\label{eq:Rp}
    \mathcal{R}:\quad x_1 \leq x_2 \leq \cdots \leq x_{N-1} \leq x_N,
\end{equation}
of the configuration space.  Where all coordinates $x_j$ are distinct, the interaction term in Hamiltonian~\eqref{eq:LLmodel} vanishes and the corresponding Schr\"odinger equation is that of a system of free particles.  Where two coordinates $x_j$ and $x_{j+1}$ coincide, the delta-function interaction potential can be recast as a boundary condition,
\begin{equation}\label{eq:interactions}
    \left[\left(\frac{\partial}{\partial x_{j+1}}-\frac{\partial}{\partial x_j}\right)-c\right]_{x_{j+1}=x_j}\psi(\{x_i\})=0,
\end{equation}
on the spatial derivatives of the wave function.  The solution then proceeds by the substitution of the unnormalized ansatz (valid on $\mathcal{R}$ only)
\begin{equation}\label{eq:BetheAnsatz}
    \psi(\{x_i\}) = \sum_{\sigma} a(\sigma) \; \mathrm{exp}\Big[i \sum_{m=1}^{N}x_m \lambda_{\sigma(m)}\Big] \; ,
\end{equation}
where $\sum_\sigma$ denotes a sum over all $N!$ permutations $\sigma=\{\sigma(j)\}$ of $\{1,2, \dots,N\}$.  Demanding that $\psi(\{x_i\})$ be an eigensolution of the Schr\"odinger equation corresponding to Hamiltonian~\eqref{eq:LLmodel} then yields the general expression 
\begin{equation}
     a(\sigma) = \prod_{k>l} \left( 1 - \frac{i c}{\lambda_{\sigma(k)}-\lambda_{\sigma(l)}} \right)
\end{equation}
for the phase factors $a(\sigma)$ that encode the effects of interactions between the particles.   The quantities $\lambda_j$ are termed the rapidities, or quasimomenta of the Bethe-ansatz wave function.  Imposing that the system be confined to a spatial domain of length $L$ and subject to periodic boundary conditions $\psi(\{x_1,\dots,x_i+L,\dots,x_N\})=\psi(\{x_1,\dots,x_i,\dots,x_N\})$ yields the set of $N$ Bethe equations~\cite{Lieb1963a,*Lieb1963b}  
\begin{equation}\label{eq:Betheeq}
    \lambda_j =\frac{2\pi}{L} m_j - \frac{2}{L} \sum_{k=1}^{N} \mathrm{arctan} \left( \frac{\lambda_j-\lambda_k}{c} \right)
\end{equation}
for the rapidities $\lambda_j$, where the ``quantum numbers'' $m_j$ are any $N$ distinct integers (half-integers) in the case that $N$ is odd (even)~\cite{Yang1969}.

Extending Eq.~\eqref{eq:BetheAnsatz} outside of the ordered sector $\mathcal{R}$ of the periodic domain using Bose symmetry, each set $\{\lambda_j\}$ of $N$ distinct rapidities obtained as a particular solution of the Bethe equations~\eqref{eq:Betheeq} defines a normalized eigenstate $|\{\lambda_j\}\rangle$ of Hamiltonian~\eqref{eq:LLmodel}, with spatial representation
\begin{align}\label{eq:eigenfunction_unordered}
    \zeta_{\{\lambda_j\}}(\{x_i\}) &\equiv \langle \{x_i\}|\{\lambda_j\}\rangle \nonumber \\
    &= A_{\{\lambda_j\}} \sum_{\sigma} \mathrm{exp}\Big[i \sum_{m=1}^{N}x_m \lambda_{\sigma(m)}\Big] \nonumber \\
    &\qquad \times \prod_{k>l}\Big(1-\frac{ic\,\mathrm{sgn}(x_k-x_l)}{\lambda_{\sigma(k)}-\lambda_{\sigma(l)}}\Big), 
\end{align}
where the normalization constant~\cite{Korepin1993} 
\begin{equation}\label{eq:norm}
    A_{\{\lambda_j\}} = \frac{   \prod_{k>l}(\lambda_k-\lambda_l)}{ \big\{ N! \;  \mathrm{det}\{M_{\{\lambda_j\}}\} \;\prod_{k>l} [(\lambda_k-\lambda_l)^2 + c^2 ] \big\}^{1/2}} \; ,
\end{equation}
with $M_{\{\lambda_j\}}$ the $N \times N$ matrix with elements
\begin{align}\label{eq:Gaudin_mtx}
    \left[M_{\{\lambda_j\}}\right]_{kl} &= \delta_{kl} \Big( L + \sum_{m=1}^{N} \frac{2c}{c^2 + (\lambda_k-\lambda_m)^2} \Big)\nonumber \\
&\qquad - \frac{2c}{c^2 + (\lambda_k-\lambda_l)^2}.
\end{align}
The set of all such eigenfunctions forms a complete orthonormal basis for (the Bose-symmetric subspace of) the $N$-particle Hilbert space on which Hamiltonian~\eqref{eq:LLmodel} acts~\cite{Dorlas1993}.  In the eigenstate $|\{\lambda_j\}\rangle$ the total energy,
\begin{equation}\label{eq:EnergyBR}
    E_{\{\lambda_j\}} = \sum_{j=1}^{N} \lambda_j^2,
\end{equation}
and total momentum,
\begin{equation}\label{eq:MomentumBR}
     P_{\{\lambda_j\}}=\sum_{j=1}^{N} \lambda_j,
\end{equation}
of the system, and indeed an infinite set of quantities $Q_{\{\lambda_j\}}^{(m)} \equiv \sum_{j=1}^N (\lambda_j)^m$ that are conserved under the action of the Hamiltonian~\eqref{eq:LLmodel}, are specified completely by the set $\{\lambda_j\}$ of rapidities that label the state.  In particular, the ground state of the system corresponds to the set of $N$ rapidities that minimize Eq.~\eqref{eq:EnergyBR} and constitute the (pseudo-)Fermi sea of the 1D Bose gas~\cite{Korepin1993}.

In this work we obtain ground- and excited-state solutions to Eq.~\eqref{eq:Betheeq} numerically using a standard Newton solver.  The numerical solution is significantly aided by the fact that the Jacobian matrix corresponding to Eq.~\eqref{eq:Betheeq} takes a simple analytical form~\cite{Yang1969}.  In practice, we exploit the fact that in the TG limit $c\to\infty$ the rapidities $\{\lambda_j\}$ are simply the single-particle momenta of a system of free spinless fermions~\cite{Korepin1993} to obtain initial guesses for the rapidities in the strongly interacting regime $c\gg 1$.  We then obtain solutions for the rapidities $\{\lambda_j\}$ at successively smaller values of $c$, providing the root-finding algorithm in each case with an initial guess for these quantities obtained from linear extrapolation of the converged solutions found at stronger interaction strengths.

\subsection{Calculation of correlation functions}\label{subsec:correlation_functions}
Throughout this article, we present results on the $m^\mathrm{th}$-order equal-time correlation functions
\begin{align}
    &G^{(m)}(x_1, \dots, x_m, x_1', \dots, x_m'; t) \nonumber \\
    &\equiv \left\langle \hat{\Psi}^\dagger (x_1) \cdots \hat{\Psi}^\dagger (x_m) \hat{\Psi}(x_1')\cdots \hat{\Psi}(x_m')\right\rangle,
\end{align}
where $\langle\cdots\rangle \equiv \mathrm{Tr}\{\hat{\rho}(t)\cdots\}$ denotes an expectation value in a Schr\"odinger-picture density matrix $\hat{\rho}(t)$, and $\hat{\Psi}^{(\dagger)}(x)$ is the annihilation (creation) operator for the Bose field.  Formally, the corresponding normalized correlation \mbox{functions} are
\begin{align*}
&g^{(m)}(x_1, \dots, x_m, x_1', \dots, x_m'; t) \nonumber \\
&\equiv \frac{G^{(m)}(x_1, \dots, x_m, x_1', \dots, x_m'; t)}{\left[\langle \hat{n}(x_1)\rangle \cdots \langle \hat{n}(x_m)\rangle \langle \hat{n}(x_1')\rangle \cdots\langle \hat{n}(x_m')\rangle\right]^{1/2}},
\end{align*}
where $\hat{n}(x) \equiv \hat{\Psi}^\dagger(x) \hat{\Psi}(x)$.   We note, however, that in the nonequilibrium scenarios we consider in this article both the initial state of the system and the Hamiltonian that generates its time evolution are translationally invariant (modulo the finite extent $L$ of the periodic geometry).  Thus, the mean density $\langle \hat{n}(x) \rangle \equiv n$ is constant in both time and space, and $g^{(m)}(x_1, \dots, x_m, x_1', \dots, x_m'; t) = G^{(m)}(x_1, \dots, x_m, x_1', \dots, x_m'; t)/n^m$.  In the remainder of this article we consider the forms of these correlation functions both in a pure (time-dependent) state $|\psi(t)\rangle$, in which case    
\begin{widetext}
\begin{align}\label{eq:Gm_pure_state}
    g^{(m)}(x_1, \dots, x_m, x_1', \dots, x_m'; t) &= \frac{1}{n^m}\langle \psi(t) | \hat{\Psi}^\dagger (x_1) \cdots \hat{\Psi}^\dagger (x_m) \hat{\Psi}(x_1')\cdots \hat{\Psi}(x_m') | \psi(t) \rangle \nonumber \\
    &=N! \! \int_{0}^{L} \! \frac{dx_{m+1} \cdots dx_N}{n^m(N-m)!} \psi^*(x_1,\dots,x_m,x_{m+1},\dots,x_N,t) \psi(x_1',\dots, x_m', x_{m+1},\dots,x_N,t)
\end{align}
and in a statistical ensemble with density matrix $\hat{\rho}_\mathrm{SE} \equiv \sum_{\{\lambda_j\}} \rho^\mathrm{SE}_{\{\lambda_j\}}|\{\lambda_j\}\rangle\langle\{\lambda_j\}|$, in which case
\begin{align}\label{eq:Gm_stat_ensemble}
    g_\mathrm{SE}^{(m)}(x_1, \dots, x_m, x_1', \dots, x_m') &= \frac{1}{n^m}\mathrm{Tr}\{\hat{\rho}_\mathrm{SE}  \hat{\Psi}^\dagger (x_1) \cdots \hat{\Psi}^\dagger (x_m) \hat{\Psi}(x_1')\cdots \hat{\Psi}(x_m')\} \nonumber \\
   &= \frac{1}{n^m}\sum_{\{\lambda_j\}} \rho^\mathrm{SE}_{\{\lambda_j\}} \langle \{ \lambda_j \} |  \hat{\Psi}^\dagger (x_1) \cdots \hat{\Psi}^\dagger (x_m) \hat{\Psi}(x_1')\cdots \hat{\Psi}(x_m') | \{ \lambda_j \} \rangle,
\end{align}
\end{widetext}
where the matrix elements of field-operator products are given in first-quantized form by Eq.~\eqref{eq:Gm_pure_state} upon replacing $\psi(\{x_i\},t) \to \zeta_{\{\lambda_j\}}(\{x_i\})$.  The evaluation of such integrals can then be performed semianalytically, following the approach of Ref.~\cite{Forrester2006}.  For this purpose, we developed a symbolic integration algorithm, which will be presented elsewhere~\cite{Zill2014}.  We note also that translational invariance of the state $|\psi(t)\rangle$ (or $\hat{\rho}_\mathrm{SE}$) also implies that the correlation functions $g^{(m)}(x_1,\dots, x_m, x_1', \dots, x_m')$ are invariant under global coordinate shifts $x\to x + d$, and thus $g^{(1)}(x,y)\equiv g^{(1)}(0,y-x)$, etc.  We focus, in particular, on the first-order correlation function $g^{(1)}(x)\equiv g^{(1)}(0,x)$, the second-order correlation function $g^{(2)}(x) \equiv g^{(2)}(0,x,x,0)$, and the local third-order coherence $g^{(3)}(0)=\langle [ \hat{\Psi}^\dagger(0) ]^3 [\hat{\Psi}(0) ]^3 \rangle/n^3$.

We note that as we work in units $\hbar=2m=1$, time (energy) has dimensions of (inverse) length squared.  Although our results depend explicitly on the number of particles $N$ in our system, the extent $L$ of our periodic geometry, and consequently the density $n\equiv N/L$ of the Bose gas, is arbitrary.  Following Ref.~\cite{Lieb1963a} we absorb the density into the dimensionless interaction-strength parameter $\gamma=c/n$.  In the thermodynamic limit $N,\,L\to\infty$ at constant $n$, the interaction strength $\gamma$ is the only parameter of the LL theory.  However, in our finite system, the particle number $N$ must also be specified.  We hereafter quote the strength of interactions in our calculations in terms of $\gamma$.  The Fermi momentum $k_F = (2\pi/L)(N-1)/2$, which is the magnitude of the largest rapidity occurring in the ground state in the TG limit~\cite{Korepin1993}, is a convenient unit of inverse length and so we often specify lengths in units of $k_F^{-1}$, energies in units of $k_F^{2}$, and times in units of $k_F^{-2}$.

\section{Dynamics following an interaction-strength quench}\label{sec:dynamics}
We now investigate the nonequilibrium dynamics of the LL model following a sudden change (quench) of the interparticle interaction strength $\gamma$.  We focus, in particular, on a quench of a system initially in the ground state $|\psi_0\rangle$ of Hamiltonian~\eqref{eq:LLmodel} in the limit of vanishing interaction strength~\cite{Gritsev2010,Muth2010a,Kormos2013,DeNardis2014a,Kormos2014,NoteA}.  We note that the corresponding spatial wave function of this initial state is simply a constant, 
\begin{equation}\label{eq:psiini}
    \psi_0(\{x_i\}) = \langle \{x_i\}| \psi_0 \rangle = L^{-N/2},
\end{equation}
and, e.g., the spatial correlation functions $g^{(1)}_{\gamma=0}(x)=1$ and $g^{(2)}_{\gamma=0}(x)=1-1/N$ in this state are also constants.  At $t=0$, we discontinuously change the interaction strength to a finite final value $\gamma>0$.  The ensuing time evolution of the state is governed by the LL Hamiltonian $\hat{H}$ [Eq.~\eqref{eq:LLmodel}] with interaction strength $\gamma$.  As $\hat{H}$ is time-independent following the quench, energy is conserved during the dynamics.  This conserved energy is the energy of the system at time $t=0^+$,  
\begin{align}\label{eq:E_final}
    E &\equiv \langle\psi(0^+)|\hat{H}|\psi(0^+)\rangle = (N-1)n^2\gamma,
\end{align}
which is easily derived by noting that the state $|\psi(0^+)\rangle$ immediately following the quench is simply the (homogeneous) prequench wave function $|\psi_0\rangle$, in which the kinetic-energy component of Hamiltonian~\eqref{eq:LLmodel} vanishes and in which the interaction energy is determined by the local second-order coherence [$g^{(2)}_{\gamma=0}(0)$] of the state.  

Formally, the time-evolving wave function is given at all times $t>0$ by 
\begin{equation}\label{eq:initialspectral}
    |\psi(t) \rangle = \sum_{\{\lambda_j\}} C_{\{\lambda_j\}}  e^{- i E_{\{\lambda_j\}} t} |\{\lambda_j\} \rangle,
\end{equation}
where the sum is over all eigenstates $|\{\lambda_j\}\rangle$ of $\hat{H}$, and the $C_{\{\lambda_j\}} \equiv \langle \{\lambda_j\} | \psi_0 \rangle$ are the overlaps between the initial state $|\psi_0\rangle$ and these eigenstates, which we calculate from their coordinate-space representations $\zeta_{\{\lambda_j\}}(\{x_i\})$~\cite{Zill2014,NoteB}.  We note, however, that only those states $|\{\lambda_j\}\rangle$ that have zero total momentum, $\sum_j \lambda_j=0$ [cf. Eq.~\eqref{eq:MomentumBR}], and are parity invariant (for which the rapidities $\{\lambda_j\}$ can be enumerated such that $\lambda_j = - \lambda_{N+1-j};\quad j=1,2,\dots,N$) have nonzero overlaps with the initial state $|\psi_0\rangle$, as discussed in Refs.~\cite{Kormos2013,Calabrese2014}. 

We primarily characterize the nonequilibrium dynamics of the system by the evolution of its equal-time correlation functions (Sec.~\ref{subsec:correlation_functions}).  These are calculated by noting that the time evolution of the expectation value of an arbitrary operator $\hat{O}$ in the time-dependent state $|\psi(t)\rangle$ is given by 
\begin{align}\label{eq:observabletime}
    \langle \hat{O} (t)\rangle &\equiv \langle \psi(t) |\hat{O}| \psi(t) \rangle  \\
    &=\!\sum_{\{\lambda_j^{}\}}\!\sum_{\{\lambda_j'\}}\! C_{\{\lambda_j'\}}^* C_{\{\lambda_j^{}\}}^{} e^{ i (E_{\{\lambda_j'\}}\! - E_{\{\lambda_j^{}\!\}}\!) t} \! \langle \{\lambda'_j\} | \hat{O} | \{\lambda_j\} \rangle. \nonumber  
\end{align}
The matrix elements $\langle \{\lambda'_j\} | \hat{O} | \{\lambda_j\} \rangle$ of observables are calculated in a similar manner to the overlaps $C_{\{\lambda_j\}}$, as we will discuss in Ref.~\cite{Zill2014}.  The computational expense incurred in evaluating these matrix elements increases exponentially with the particle number $N$, placing a strong practical constraint on the system sizes we can describe with our coordinate Bethe-ansatz approach.  In the remainder of this article, unless otherwise specified, we always consider a quench of $N=5$ particles.

Assuming that all energies $E_{\{\lambda_j\}}$ of the contributing eigenstates $|\{\lambda_j\}\rangle$ are nondegenerate, the (infinite-)time average of Eq.~\eqref{eq:observabletime} is 
\begin{align}\label{eq:DE}
    \langle \hat{O} \rangle_{\mathrm{DE}} &= \lim_{\tau\to\infty} \frac{1}{\tau} \int_0^\tau\! dt\, \langle\psi(t)|\hat{O}|\psi(t)\rangle \nonumber \\
    &= \sum_{\{\lambda_j\}}  |C_{\{\lambda_j\}}|^2 \langle \{\lambda_j\} | \hat{O} | \{\lambda_j\} \rangle, 
\end{align}
which we identify as the expectation value of $\hat{O}$ in the density matrix,
\begin{align}\label{eq:DEdmt}
    \hat{\rho}_\mathrm{DE} &= \sum_{\{\lambda_j\}} |C_{\{\lambda_j\}}|^2 |\{\lambda_j\}\rangle \langle \{\lambda_j\}|,
\end{align}
of the DE~\cite{Rigol2008}.  A finite system such as we consider here does not exhibit true relaxation, in which the instantaneous density matrix of the system (and therefore all observables) becomes stationary in the long-time limit $t\to\infty$, but will instead exhibit recurrences~\cite{Bocchieri1957,Schulman1978}.  However, the dephasing of the energy eigenstates is expected to lead, quite generically, to observables fluctuating about reasonably well-defined mean values consistent with the DE predictions~\cite{Rigol2008}.  Numerical results for a number of systems indicate that the relative magnitude of these fluctuations scales towards zero with increasing system size and thus that observables relax to the predictions of the DE in the thermodynamic limit (see, e.g., Refs.~\cite{Rigol2009a,Rigol2009b,Cassidy2011}).  Establishing whether the LL system relaxes to the DE following an interaction-strength quench in the thermodynamic limit is beyond the scope of this article.  We therefore simply regard the DE defined by Eq.~\eqref{eq:DEdmt} as the ensemble appropriate to describe the relaxed state of our finite-sized system. 

We note that formally the sums in Eqs.~\eqref{eq:initialspectral}--\eqref{eq:DEdmt} range over an infinite number of LL eigenstates.  In practice, we include only a finite number of eigenstates in our calculations and thus truncate the sums in Eqs.~\eqref{eq:initialspectral}--\eqref{eq:DEdmt}.  As we discuss in Appendix~\ref{app:cutoff}, we retain all eigenstates $|\{\lambda_j\}\rangle$ that have (absolute) overlap with $|\psi_0\rangle$ greater than some threshold value.  The accuracy of our results can then be quantified by considering the saturation of the sum rules associated with the normalization (cf. Ref.~\cite{Mossel2010}) and energy of the wave function $|\psi(t)\rangle$ (see Appendix~\ref{app:cutoff}).

\subsection{First-order correlations}\label{subsec:noneqfirst} 
We begin our characterization of the nonequilibrium dynamics of the LL system following the quench by considering the first-order (or one-body) correlations of the system.  As the translational invariance of the initial state $|\psi_0\rangle$ is preserved under the evolution generated by $\hat{H}$, the first-order correlations are at all times completely described by the momentum distribution  
\begin{equation}\label{eq:mom_dist}
    \widetilde{n}(k,t) = n \int_{0}^{L} dx \; e^{-i k x} g^{(1)}(x,t).
\end{equation}
We note that, in our finite periodic geometry, the single-particle momentum $k$ is quantized and takes discrete values $k_j=2\pi j / L$, where $j$ is an integer.  In the initial state, all particles occupy the ground (zero-momentum) single-particle orbital [i.e., $\widetilde{n}(0,t=0^-)\equiv N$], and at times $t>0$ the presence of finite interparticle interactions $\gamma>0$ induces partial redistribution of this population over single-particle modes with finite momenta $|k|>0$.  The ensuing dynamics of the momentum distribution have previously been considered in the nonequilibrium field-theoretical studies of the dynamics of the LL model presented in Refs.~\cite{Gasenzer2005,Berges2007,Branschadel2008,Gasenzer2008}, whereas in later works the focus has been set primarily on the second-order (density-density) correlations~\cite{Gritsev2010,Muth2010a,Mossel2012a,DeNardis2014a}.  Exceptions can be found in Refs.~\cite{Kormos2013,Kormos2014}, which presented results for $g^{(1)}(x)$ in the stationary state following a quench to the TG limit (in which case the Bose-Fermi mapping and Wick's theorem can be used to simplify the calculation significantly) and in Ref.~\cite{DeNardis2014b}, which details the calculation of the dynamical evolution of $g^{(1)}(x,t)$ in the same TG-limit quench scenario.
~\\

\begin{figure}
    \includegraphics[width=0.48\textwidth]{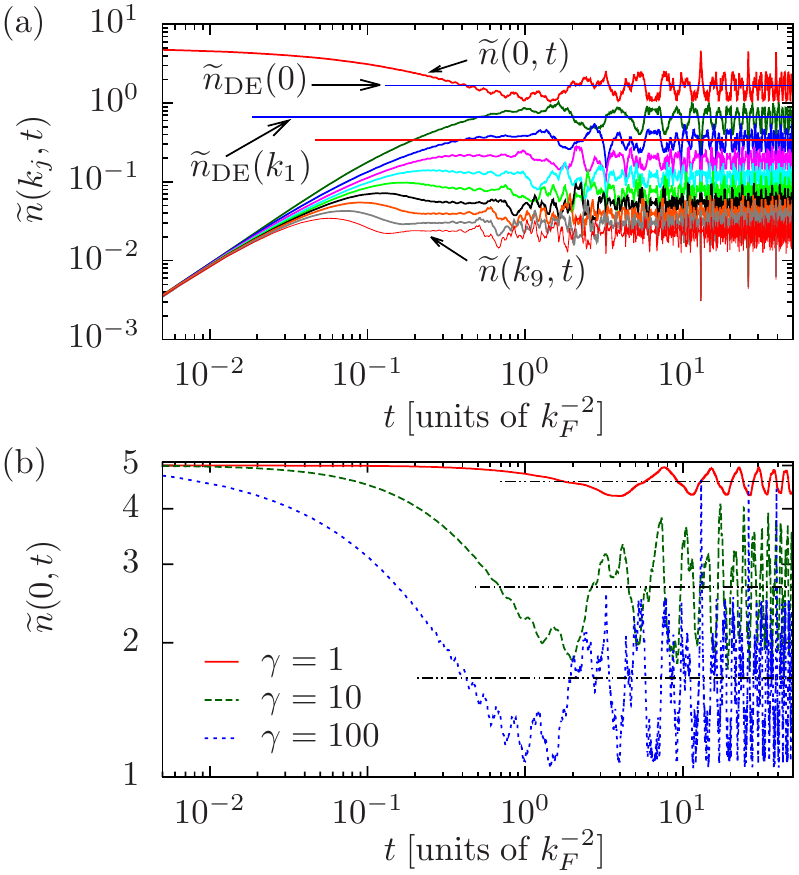}
    \caption{\label{fig:modes} (Color online)  (a) Time evolution of the occupations of the first ten non-negative momentum modes, $\widetilde{n}(k_j,t) \; (j=0, 1, \dots, 9)$, for $N=5$ particles following a quench of the interaction strength from zero to $\gamma=100$.  Horizontal solid lines indicate the equilibrium values $\widetilde{n}_\mathrm{DE}(k_j)$ predicted by the DE, for the first three non-negative momentum modes.  (b) Time evolution of the zero-momentum occupation $\widetilde{n}(0,t)$ following quenches of $N=5$ particles to $\gamma = 1, \; 10,$ and $100$.  The horizontal dot-dashed lines indicate the corresponding DE values $\widetilde{n}_\mathrm{DE}(0)$.} 
\end{figure}

In Fig.~\ref{fig:modes}(a) we plot the evolution of the occupations of the first ten non-negative momentum modes, $\widetilde{n}(k_j,t)\; (j=0,1,\dots, 9)$, following a quench to $\gamma=100$.  In the limit $t\to 0^+$, the occupations of all nonzero momentum modes rise at a common $k$-independent rate, due to the purely local nature of the delta-function interaction potential, which corresponds to a momentum-independent coupling~\cite{Berges2007}.  As time progresses, the zero-momentum occupation $\widetilde{n}(0,t)$ correspondingly decreases, and the occupation of each nonzero momentum mode $k_j$ levels off and fluctuates about its DE value $\widetilde{n}_\mathrm{DE}(k_j)$ [see Eq.~\eqref{eq:DE}], which we indicate in Fig.~\ref{fig:modes}(a) for the first three non-negative momenta $k_j \; (j=0,1,2)$ (horizontal solid lines).  The time evolution of the momentum distribution shown in Fig.~\ref{fig:modes}(a) is similar to the results obtained with functional-integral field-theory methods~\cite{Gasenzer2005,Berges2007,Branschadel2008,Gasenzer2008}.  In particular, the populations of higher momentum modes stop increasing and settle to their DE values (about which they fluctuate) more rapidly than those of lower momentum modes, indicating that nonlocal first-order correlations relax increasingly rapidly on decreasing length scales (cf., e.g., Refs.~\cite{Langen2013,Gasenzer2005,Berges2007,Branschadel2008,Sykes2014}).  We note, however, that the momentum distribution here, similarly to that observed for a quench to the strongly interacting regime in Ref.~\cite{Gasenzer2008}, appears to evolve directly to a stationary state, without exhibiting any intermediary period of quasistationary relaxation such as that observed for quenches to weak interaction strengths in Refs.~\cite{Gasenzer2005,Berges2007,Branschadel2008}.  
~\\
 
Qualitatively similar evolution is observed for any value of the final interaction strength $\gamma$, but both the form of the DE momentum distribution $\widetilde{n}_\mathrm{DE}(k_j)$ and the time scales on which mode occupancies reach their DE values depend strongly on $\gamma$.  A useful summary statistic by which to compare the relaxation of first-order correlations between quenches is the occupation $\widetilde{n}(0,t)$ of the zero-momentum mode, the dynamical evolution of which we plot in Fig.~\ref{fig:modes}(b) for $\gamma=1, \; 10,$ and $100$.  We note that in the case $\gamma = 1$, $\widetilde{n}(0,t)$ exhibits near-monochromatic oscillations over time.   For a larger interaction strength $\gamma=10$, the zero-momentum occupation $\widetilde{n}(0,t)$ first crosses $\widetilde{n}_\mathrm{DE}(0)$ earlier (at time $t\approx 0.7\,k_F^{-2}$), after which it exhibits less regular, more intricately structured fluctuations about $\widetilde{n}_\mathrm{DE}(0)$.  In the quench to the Tonks regime ($\gamma=100$), the DE value is first reached even earlier (at time $t\approx 0.4\,k_F^{-2}$), and we note also that the fluctuations of $\widetilde{n}(0,t)$ around $\widetilde{n}_\mathrm{DE}(0)$ are, in general, somewhat smaller than those observed in the quench to $\gamma=10$, although in this case $\widetilde{n}(0,t)$ also exhibits near-complete revival peaks, in which it returns close to its initial value.     
~\\

\subsection{Second-order correlations}\label{subsec:noneqsecond}
\begin{figure}
    \includegraphics[width=0.48\textwidth]{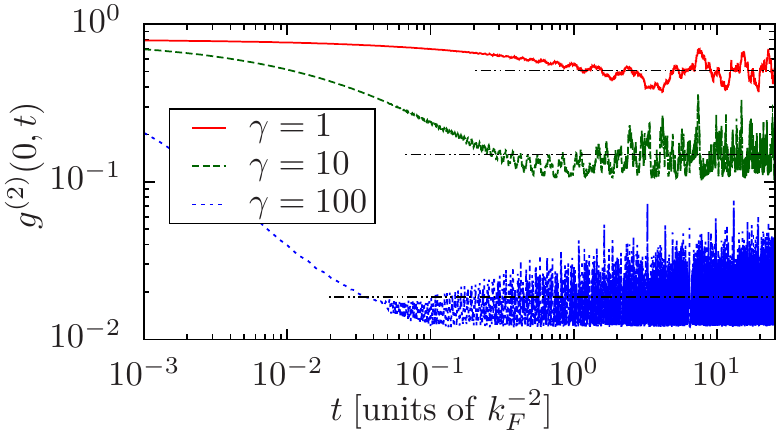}
    \caption{\label{fig:g2xzero} (Color online)  Time evolution of the local second-order coherence $g^{(2)}(0,t)$ following quenches of the interaction strength to $\gamma = 1, 10,$~and~$100$ for $N=5$ particles. Horizontal dot-dashed lines indicate the corresponding equilibrium values $g^{(2)}_\mathrm{DE}(0)$ predicted by the DE.}
\end{figure}

We now extend our characterization of the relaxation dynamics of the LL system to the second-order (or two-body) correlations of the Bose field.  We focus first on the local second-order coherence $g^{(2)}(0,t)$, the time evolution of which we plot in Fig.~\ref{fig:g2xzero} for $\gamma=1, 10$,~and~$100$. 
Similarly to $\widetilde{n}(0,t)$, as time evolves the local second-order coherence decays from its initial value $g^{(2)}(0,t=0)=1-N^{-1}$ before settling down to fluctuate about the prediction $g^{(2)}_\mathrm{DE}(0)$ of the diagonal ensemble.  In the case $\gamma=1$, $g^{(2)}(0,t)$ decays over a time scale similar to that over which the corresponding zero-momentum occupation $\widetilde{n}(0,t)$ decays and subsequently exhibits similar near-regular oscillations about its DE value $g^{(2)}_\mathrm{DE}(0)$ [cf.~Fig.~\ref{fig:modes}(b)].  As the final interaction strength $\gamma$ increases, $g^{(2)}(0,t)$ reaches its time-averaged value $g^{(2)}_\mathrm{DE}(0)$ increasingly rapidly, and this value itself decreases.  We note that although this behavior is qualitatively consistent with that observed for the zero-momentum occupation in Fig.~\ref{fig:modes}(b), at large final interaction strengths $g^{(2)}(0,t)$ decays to its DE value much more rapidly than the nonlocal quantity $\widetilde{n}(0,t)$.  
~\\

\begin{figure}
    \includegraphics[width=0.48\textwidth]{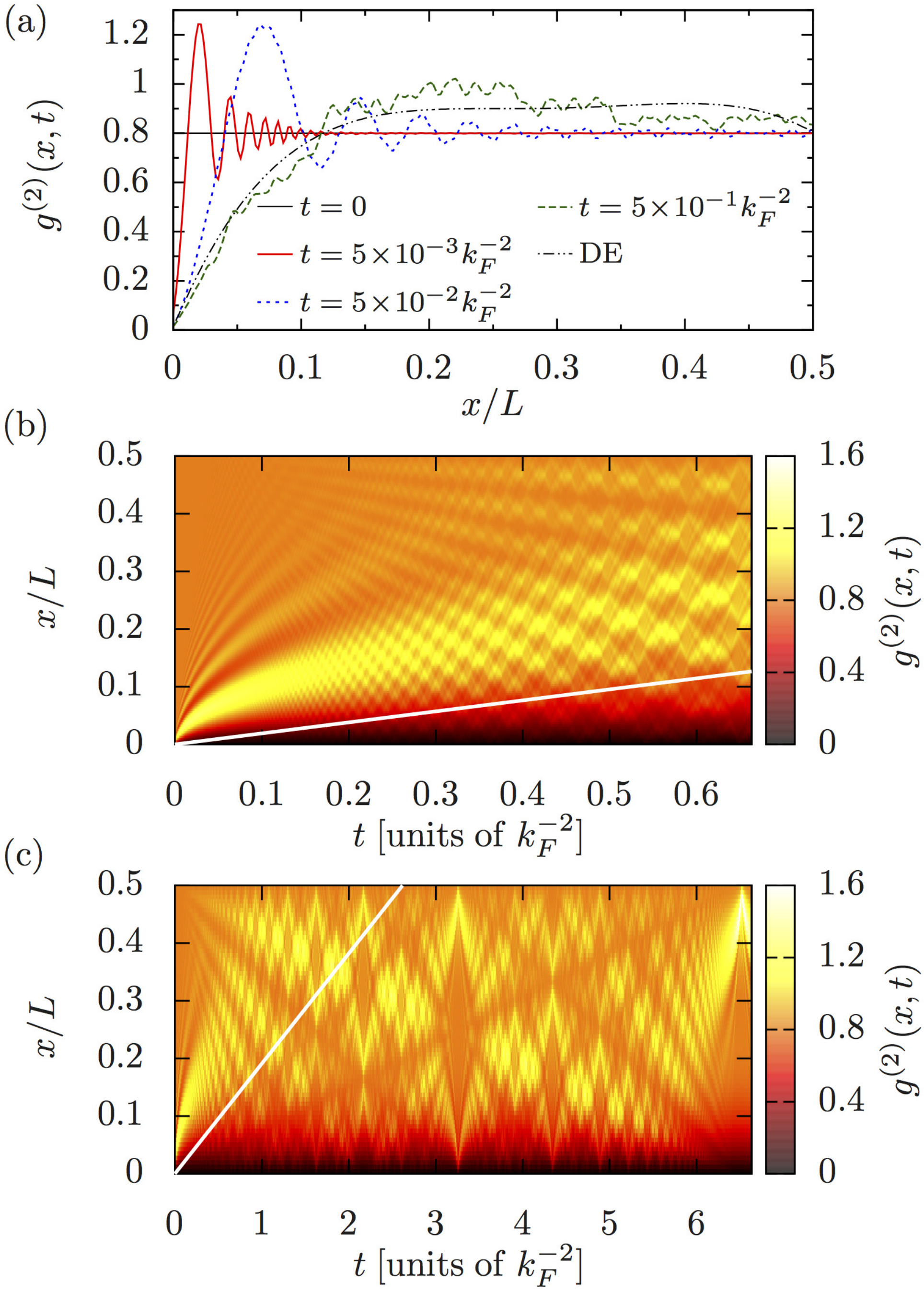}
    \caption{\label{fig:g2x} (Color online)  Time evolution of the nonlocal second-order coherence $g^{(2)}(x,t)$ following a quench of $N=5$ particles to $\gamma=100$. (a) Correlation function $g^{(2)}(x)$ at four representative times.  The black dot-dashed line indicates the prediction of the DE for the equilibrium form of this function.  (b) Evolution of $g^{(2)}(x,t)$ for short times $t\leq \pi/5\, k_F^{-2}$ and (c) longer times $t\leq 2\pi\, k_F^{-2}$.  The white solid lines in (b) and (c) indicate the trajectory $x=v_s t$ of a particle propagating away from the origin at the zero-temperature speed of sound $v_s$ of the LL system with interaction strength $\gamma=100$ (see text).}
\end{figure}

In Fig.~\ref{fig:g2x} we present the time evolution of the full nonlocal second-order correlation function $g^{(2)}(x,t)$ for a quench to $\gamma=100$.
Figure~\ref{fig:g2x}(a) shows the dependence of this function on the separation $x$ at four representative times.  At time $t=0$, $g^{(2)}(x)$ has the $x$-independent form appropriate to the noninteracting ground state (black horizontal line).  By time $t=5\times 10^{-3}\,k_F^{-2}$ (red solid line) the local second-order coherence $g^{(2)}(0,t)$ has decreased to $\approx 7\times 10^{-2}$, and $g^{(2)}(x,t)$ exhibits a maximum at a finite spatial separation and a decaying oscillatory structure past this maximum.  The appearance of such an \emph{increase} in $g^{(2)}(x,t)$ at some finite $x$ is required by conservation of the integrated second-order correlation function $\int_0^L\!dx\, g^{(2)}(x,t)$ (which itself follows from conservation of particle number and total momentum during the evolution)~\cite{Muth2010a}.  By time $t=5\times 10^{-2}\,k_F^{-2}$ (blue dotted line) the maximum in $g^{(2)}(x,t)$ and the smaller subsidiary maxima and minima that accompany it have propagated to larger separations.  The oscillations in $g^{(2)}(x)$ appear quite distorted at time $t=5\times 10^{-1}\,k_F^{-2}$ (green dashed line), though the broad envelope of this function is at this time comparable to the DE prediction for the equilibrium form of $g^{(2)}(x)$ (black dot-dashed line).  The formation and propagation of such a ``correlation wave'' was previously observed in phase-space~\cite{Deuar2006} and matrix-product-state~\cite{Muth2010a} simulations of quenches from zero to finite $\gamma$ within a Bose-Hubbard lattice discretization of the LL model and in Bethe-ansatz-based simulations of a quench of the continuous gas to the TG limit $\gamma\to\infty$~\cite{Gritsev2010,NoteC}.  

Figure~\ref{fig:g2x}(b) gives a more complete picture of the evolution of $g^{(2)}(x,t)$ following the quench.  We observe that the oscillations in this function initially propagate rapidly, but then slow and disperse as time progresses.  By time $t=0.6\,k_F^{-2}$ the primary maximum of $g^{(2)}(x,t)$ has dispersed to a width comparable to $L/2$, though additional modulations, due to interference between oscillations propagating in opposite directions around the periodic geometry, have by this time destroyed any meaningful distinction between the (initially well-resolved) individual maxima and minima of the correlation wave.  Nevertheless, the behavior of $g^{(2)}(x,t)$ at early times $t\lesssim 0.5\,k_F^{-2}$ is consistent with analytical results for a quench to the TG limit recently obtained in Ref.~\cite{Kormos2014}, which found that the maxima of the correlation wave propagate with an algebraically decaying velocity $v \propto 1/\sqrt{t}$.  On longer time scales [Fig.~\ref{fig:g2x}(c)] $g^{(2)}(x,t)$ exhibits a more complicated structure.  In particular, $g^{(2)}(x,t)$ appears crisscrossed by a number of solitonlike ``density'' dips.  The slowest of these propagates at approximately $40\%$ of the speed of sound $v_s = 2\pi (1-4/\gamma) N/L = 2.4\,k_F$~\cite{Lieb1963a,*Lieb1963b,Cazalilla2004,NoteD} of a zero-temperature system with interaction strength $\gamma=100$ [indicated by white solid lines in Figs.~\ref{fig:g2x}(b)~and~\ref{fig:g2x}(c)].  This slowest-moving dip is accompanied by similar depressions propagating at integer multiples of its velocity---although the more rapidly moving dips are less well resolved in Fig.~\ref{fig:g2x}(c).  We discuss the significance of this particular set of velocities further in Sec.~\ref{subsec:Fidelity}. 

\begin{figure}
    \includegraphics[width=0.48\textwidth]{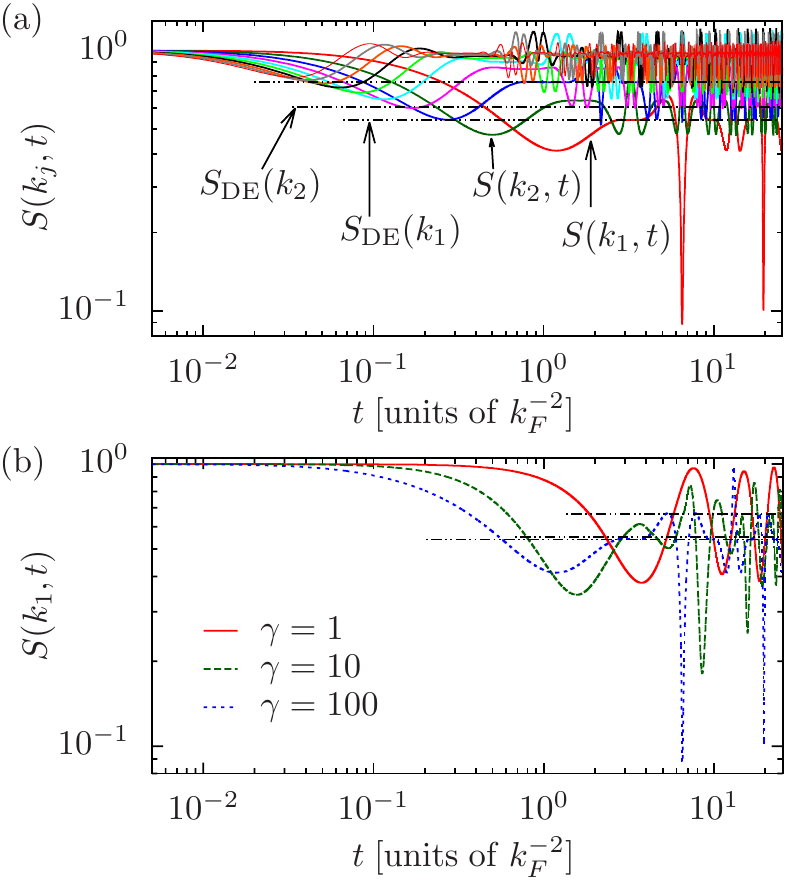}
    \caption{\label{fig:structfactdyn} (Color online)  Time evolution of the structure factor for $N=5$ particles.  (a) Components of the structure factor at the first ten positive momenta, $S(k_j,t)\; (j=1,2,\dots,10)$, for a quench to $\gamma=100$.  Horizontal dot-dashed lines indicate the DE values $S_\mathrm{DE}(k_j)$ for $j=1,2,$~and~$3$ (bottom to top). (b) First positive-momentum component $S(k_1,t)$ of the structure factor, for $\gamma=1, 10,$~and~$100$.  Horizontal dot-dashed lines indicate the DE values $S_\mathrm{DE}(k_1)$ for $\gamma=1, 10,$~and~$100$ (top to bottom).}
\end{figure}

We now consider an alternative characterization of the time development of second-order correlations in the system, given by the instantaneous structure factor~\cite{Pitaevskii2003}
\begin{equation}\label{eq:structfact}
    S(k,t) = 1 + n \int_{0}^{L} dx \; e^{-i k x} \left[g^{(2)}(x,t) - 1\right].
\end{equation}
We note that particle-number conservation and translational invariance imply that $S(0,t)=0$ at all times $t$.  In Fig.~\ref{fig:structfactdyn}(a) we therefore plot the time development of the structure factor, evaluated at the first ten positive wave vectors $k_j\; (j=1,2,\dots,10)$ in our finite periodic geometry, for a quench to $\gamma=100$.

We note that the behavior of the individual components $S(k_j,t)$ of the structure factor is opposite to that of the occupations $\widetilde{n}(k_j,t)$ of nonzero momentum modes $k_j$ for this quench [Fig.~\ref{fig:modes}(a)], in that the $S(k_j,t)$ begin at unity and decay towards their DE values $S_\mathrm{DE}(k_j,t)$ as time progresses.  Moreover, in contrast to the momentum occupations $\widetilde{n}(k_j,t)\;(j>0)$, which initially rise uniformly, the components $S(k_j,t)$ of the structure factor at distinct momenta $k_j$ decay at distinct rates even in the limit $t \to 0^+$.  However, just as observed for the momentum distribution, components of the structure factor at higher momenta reach their first turning points and settle (with large fluctuations) around their DE values more rapidly than those components at lower momenta.  In particular, $S(k_1,t)$ is the last component to reach its turning point and, in general, fluctuates more slowly about its time-averaged value $S_\mathrm{DE}(k_1)$ than higher-momentum components, although its oscillations include large excursions towards zero and unity.  This can be seen more clearly in Fig.~\ref{fig:structfactdyn}(b), where we compare the time evolution of $S(k_1,t)$ (which we take as a simple summary measure for the evolution of the structure factor) for quenches to $\gamma=1,10,$~and~$100$.  Similarly to $\widetilde{n}(0,t)$, the structure-factor component $S(k_1,t)$ exhibits approximately monochromatic oscillations for the quench to $\gamma=1$.  Moreover, $S(k_1,t)$ first crosses its DE value sooner, and exhibits progressively less-regular oscillations, with increasing $\gamma$.  We observe that for $\gamma=100$, the component $S(k_1,t)$ exhibits a large fluctuation towards zero at time $t\approx 6.51\,k_F^{-2}$.  Considering Fig.~\ref{fig:g2x}(c), we see that this time also corresponds to that at which the solitonlike correlation dip in $g^{(2)}(x,t)$ that emerges following the quench, propagating at a velocity $\approx1.0\,k_F$, reaches $x=L/2$.  A large fluctuation of $S(k_1,t)$ to a value close to unity occurs at time $t\approx 13.1\,k_F^{-2}$, coinciding with the quasirecurrence of $\widetilde{n}(0,t)$ in Fig.~\ref{fig:modes}(a), and a second fluctuation of $S(k_1,t)$ towards zero (somewhat smaller than the first) occurs at time $t\approx 19.9\,k_F^{-2}$, indicating a (quasi-)regular pattern of large fluctuations in the correlations of the system.

\subsection{Fidelity}\label{subsec:Fidelity} 
So far our characterizations of the nonequilibrium dynamics of the LL model have considered only the one- and two-body correlations of the system.  We now consider a quantity that allows us to characterize the relaxation of the system in the $N$-body state space of the LL model: the quantum fidelity~\cite{Nielsen2000}.  The fidelity provides a measure of ``closeness'' between two quantum states and, when evaluated between a pure state $|\chi\rangle$ and an arbitrary (pure or mixed) density matrix $\hat{\sigma}$, takes the form $F(|\chi\rangle,\hat{\sigma})=\langle \chi|\hat{\sigma}|\chi\rangle$.  We note first that the fidelity
\begin{align}\label{eq:F_DE}
    F_\mathrm{DE} &= \langle \psi(t) | \hat{\rho}_\mathrm{DE} | \psi(t) \rangle  = \sum_{\{\lambda_j\}} |C_{\{\lambda_j\}}|^4 
\end{align}
between the time-evolving state $|\psi(t)\rangle$ and the DE density matrix is time independent, as $\hat{\rho}_\mathrm{DE}$ is (by definition) diagonal in the energy eigenbasis of $\hat{H}$ and therefore invariant under the action of the time-displacement operator $\hat{U}(t)=\sum_{\{\lambda_j\}} \exp(-i E_{\{\lambda_j\}} t)|\{\lambda_j\}\rangle\langle \{\lambda_j\}|$.  In fact, the fidelity $F_\mathrm{DE}$ is simply the inverse participation ratio (IPR)~\cite{Zelevinsky1996} of the initial state $|\psi_0\rangle$ in the energy eigenbasis of $\hat{H}$.   

We characterize the dynamics of the time-evolving state vector $|\psi(t)\rangle$ in the $N$-body Hilbert space by the fidelity between $|\psi(t)\rangle$ and the initial state $|\psi_0\rangle$ of the system:   
\begin{align}\label{eq:fidelity}
    F(t) &= | \langle \psi_0 | \psi(t) \rangle |^2 \nonumber \\
    &= \sum_{\{\lambda_j\}} \sum_{\{\lambda'_j\}} |C_{\{\lambda_j\}}|^2 |C_{\{\lambda'_j\}}|^2 e^{i (E_{\{\lambda_j\}} - E_{\{\lambda'_j\}}\!) t}. 
\end{align}
This quantity provides a characterization of the dephasing of energy eigenstates that underlies the relaxation of the system to the DE~\cite{Rigol2008}. We note in particular that, in the absence of degeneracies in the energy spectrum, the time average of the fidelity $\lim_{\tau\to\infty} (1/\tau)\int_0^\tau\!dt\,F(t) = F_\mathrm{DE}$ (see, e.g., Ref.~\cite{Torres-Herrera2014} and references therein).   

\begin{figure}
    \includegraphics[width=0.48\textwidth]{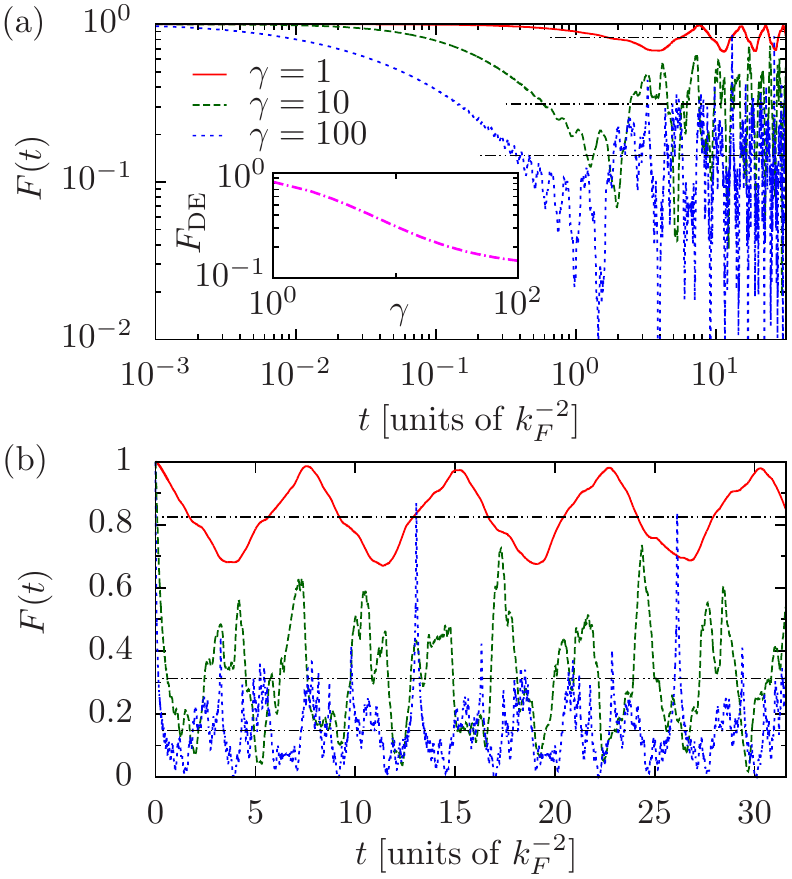}
    \caption{\label{fig:fidelity} (Color online)  (a) Fidelity $F(t)$ between time-evolving state $|\psi(t)\rangle$ and initial state $|\psi_0\rangle$.  Horizontal dot-dashed lines indicate the corresponding DE values $F_\mathrm{DE}$.  Inset: Fidelity $F_\mathrm{DE}$ between DE density matrix $\hat{\rho}_\mathrm{DE}$ and initial state $|\psi_0\rangle$ (i.e., IPR of $|\psi_0\rangle$ in the eigenstates of $\hat{H}$) as a function of $\gamma$.  (b) The same data as (a) on a linear scale.}
\end{figure}

In Fig.~\ref{fig:fidelity}(a) we plot the fidelity $F(t)$ as a function of time for $N=5$ particles and final interaction strengths $\gamma=1, 10,$~and~$100$.  
We observe that for each value of $\gamma$, the evolution of $F(t)$ is qualitatively similar to the corresponding evolution of the zero-momentum occupation $\widetilde{n}(0,t)$ [Fig.~\ref{fig:modes}(b)].  For the quench to $\gamma=1$, the fidelity exhibits near-monochromatic oscillations around its DE value.  We observe that for this quench, the IPR $F_\mathrm{DE}\approx 0.83$, implying that few eigenstates contribute significantly to the DE (note that $F_\mathrm{DE}\to1$ in the limit that $\hat{\rho}_\mathrm{DE}$ is pure).  In fact, for the quench to $\gamma=1$, the two most highly occupied energy eigenstates, with populations $n^{(0)}=|C^{(0)}|^2\approx 0.903$ and $n^{(1)}=|C^{(1)}|^2\approx0.073$, account for the majority of the norm of $|\psi(t)\rangle$, with more highly excited states accounting for the remaining $\approx 2.5\%$.  Thus, the postquench system can be regarded to a good approximation as a superposition of the ground state and the lowest-lying excited state that has finite overlap with $|\psi_0\rangle$, yielding a monochromatic oscillation in $F(t)$ with a period $t_1=2\pi/(E^{(1)}-E^{(0)}) \approx 7.52\,k_F^{-2}$, which indeed appears consistent with the primary frequency component of $F(t)$ for this quench.   This behavior is straightforward to understand, as the finite extent of the system induces a finite-size gap in the excitation spectrum.  As we discuss in Appendix~\ref{app:energy_and_finite_size_gap}, this gap strongly suppresses the excitation of the system in quenches to small values of $\gamma$, yielding effectively two-level dynamics~\cite{NoteE}.    

As the final interaction strength $\gamma$ increases, the IPR $F_\mathrm{DE}$ of $|\psi_0\rangle$ in the eigenstates of $\hat{H}$ decreases significantly [inset to Fig.~\ref{fig:fidelity}(a)].  For $\gamma=10$, we find $F_\mathrm{DE}\approx 0.31$, and in this case $F(t)$ is a strongly irregular function, composed of many frequency components, and more clearly exhibits a rapid initial decay [see the linear plot of $F(t)$ in Fig.~\ref{fig:fidelity}(b)], followed by (large) fluctuations about its temporal mean $F_\mathrm{DE}$.  We note that this decay of $F(t)$ towards $F_\mathrm{DE}$ has a simple physical interpretation.  As $F_\mathrm{DE}$ is the average of the fidelities between $|\psi(t)\rangle$ and the eigenstates $|\{\lambda_j\}\rangle$ of $\hat{H}$, weighted by their populations in $\hat{\rho}_\mathrm{DE}$, when $F(t)=F_\mathrm{DE}$ the state $|\psi(t)\rangle$ is equally close to $|\psi_0\rangle$ as it is to a typical state in the DE, indicating a loss of ``memory'' of the initial state.   

For $\gamma=100$, the IPR ($F_\mathrm{DE}\approx0.15$) and the typical magnitude of the fluctuations of $F(t)$ about it are again smaller than for $\gamma=10$.  Moreover, the evolution of $F(t)$ appears even more irregular in this case.  However, although the typical fluctuations of $F(t)$ are comparatively small, we note that $F(t)$ also exhibits sharp, sudden fluctuations towards values $\approx0.8$, and indeed closer to unity than the largest fluctuations exhibited by $F(t)$ for $\gamma=10$.  We identify the appearance of these quasirecurrences as resulting from the proximity of the system to the TG limit $\gamma\to\infty$~\cite{Kaminishi2013}.  As $\gamma$ is increased towards the TG limit, the spectrum of $\hat{H}$ approaches that of free fermions in the periodic ring geometry, which yields perfect recurrences of the initial state on comparatively short time scales, due to the commensurability of eigenstate energies.  In particular, in the TG limit the energies of eigenstates contributing to the DE are all integer multiples of $\delta \varepsilon=2k_1^2\equiv 8\pi^2/L^2$ (where the factor of $2$ is due to the restriction to parity-invariant eigenstates), yielding a recurrence time $t_r^{\mathrm{(TG)}} = 2 \pi / \delta \varepsilon = L^2/4\pi$.  For the quenches we consider here with $N=5$, the Fermi momentum $k_F=4\pi/L$, and thus $t_r^{\mathrm{(TG)}} = 4\pi\,k_F^{-2}$.  We therefore expect the sharp quasirevival evident in $F(t)$ at $t\approx 13.1\,k_F^{-2}$ to shift to earlier times and increase in magnitude as $\gamma$ is increased, ultimately becoming a perfect recurrence $[F(t_r^{\mathrm{(TG)}})=1$] in the TG limit~\cite{NoteF}.  This insight also helps us to understand the appearance of the solitonic dip in $g^{(2)}(x,t)$ [Fig.~\ref{fig:g2x}(c)] traveling at $\approx 40\%$ of the speed of sound $v_s$: Complete recurrence of the system at time $t_r^{\mathrm{(TG)}}$ would imply a minimum speed $v_\mathrm{min} = L / t_r^{\mathrm{(TG)}}$ that any (persistent) disturbance in the nonlocal correlation functions of the system can travel at, in order that it returns to its starting position when the recurrence occurs.  For $N=5$ the minimum velocity $v_\mathrm{min}=k_F$, whereas the Fermi velocity and speed of sound (in the TG limit) $v_F=2.5\,k_F$~\cite{NoteD}.  We therefore interpret the slow-moving density depression in Fig.~\ref{fig:g2x}(c) as a precursor to a solitonic disturbance propagating at $v_\mathrm{min}$ in the TG limit and the more rapidly moving dips as traveling at integer multiples of this velocity~\cite{NoteG}.  We note also that as the thermodynamic limit is approached (i.e., increasing $N$ at fixed density), the recurrence time diverges like $N^2$ and the minimum velocity vanishes like $1/N$; i.e., the discrete spectrum of permitted velocities becomes a continuum.

\subsection{Relaxation time scales}
\begin{figure}
    \includegraphics[width=0.48\textwidth]{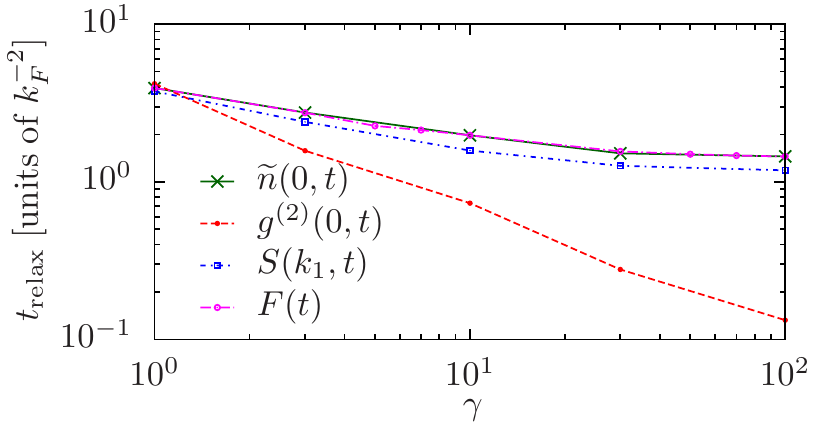}
    \caption{\label{fig:times} (Color online) Relaxation time scales (defined by the first crossing of the DE value; see text) for the zero-momentum occupation $\widetilde{n}(0,t)$, local second-order coherence $g^{(2)}(0,t)$, first nonzero momentum component $S(k_1,t)$ of the instantaneous structure factor, and fidelity $F(t)$, for a quench of $N=5$ particles.}
\end{figure}

Our results for first- and second-order correlations of the LL system following the quench, together with the fidelity $F(t)$ between the state at time $t$ and the initial state, indicate that our finite-size calculations exhibit behavior consistent with the notion of relaxation of a quantum system due to the dephasing of energy eigenstates~\cite{Rigol2008}, at least for large final interaction strengths $\gamma\gg 1$.  Here we consider the dependence of the time scales over which these quantities relax on $\gamma$.  We note that in our finite-size calculations, quantities do not, in general, show decay over sufficiently long time scales that particular functional forms (such as exponential or power-law decay) can be fitted to extract relaxation rates (or exponents).  We therefore simply associate, with each quantity we consider, a relaxation time defined as the time at which that quantity first reaches its time-averaged (DE) value.  In this manner we extract from the results of our calculations relaxation times $t_\mathrm{relax}$ for the zero-momentum occupation $\widetilde{n}(0,t)$, local second-order coherence $g^{(2)}(0,t)$, structure-factor component $S(k_1,t)$, and fidelity $F(t)$.  We plot these relaxation times $t_\mathrm{relax}$ as functions of the final interaction strength $\gamma$ in Fig.~\ref{fig:times}.  

It is clear from this figure that (as we have noted in Sec.~\ref{subsec:noneqsecond}) the local second-order coherence $g^{(2)}(0,t)$ relaxes much more quickly than $\widetilde{n}(0,t)$, aside from the strongly finite-size limited case $\gamma=1$.  Moreover, the relaxation time for the local quantity $g^{(2)}(0,t)$ decreases steadily with increasing $\gamma$ (consistent with the results of Ref.~\cite{Muth2010a}), whereas the relaxation time for the nonlocal quantity $\widetilde{n}(0,t)$ appears to saturate to a limiting value $\sim 1.5\, k_F^{-2}$ as $\gamma\to\infty$.  We note also that the relaxation time of the fidelity $F(t)$ is essentially equal to that of $\widetilde{n}(0,t)$ at each $\gamma$.  The relaxation time of $S(k_1,t)$ is, for each value of $\gamma$, somewhat smaller than that of $F(t)$ and $\widetilde{n}(0,t)$, though inspection of Fig.~\ref{fig:structfactdyn} suggests that this discrepancy arises due to the functional form of $S(k_1,t)$, which is perhaps not ideally suited to our particular definition of $t_\mathrm{relax}$.  

As the decay of the fidelity $F(t)$ quantifies the dephasing of the energy eigenstates $|\{\lambda_j\}\rangle$ of the system, we regard its evolution as the fundamental characterization of relaxation in our unitarily evolving system.  Our results here indicate that the relaxation of nonlocal quantities such as $\widetilde{n}(0,t)$~and~$S(k_1,t)$ is directly associated with the relaxation of $F(t)$ and that these experimentally relevant quantities serve as effective probes of the relaxation of the $N$-particle quantum system as a whole.  Finally in this section, we note that, on general principles, the time taken for $\widetilde{n}(0,t)$ to relax to its DE value should diverge with the time taken for correlations to traverse the system extent, which is $\propto N$ at fixed density $n$.  This should be contrasted with both the $\propto N^2$ scaling of the \mbox{(quasi-)}recurrence time scale and the essentially system-size-independent time scale for the relaxation of $g^{(2)}(0,t)$, which is determined by local physical mechanisms~\cite{Muth2010a}.

\section{Comparison of relaxed state to thermal equilibrium}\label{sec:StatMech}  
In this section we compare the correlations of the relaxed state of the system described by the DE with those that would be obtained if, following the quench, the system relaxed to thermal equilibrium.  Construction of the microcanonical ensemble is hampered by the small system size, combined with the sparse spectrum of the integrable LL Hamiltonian~\eqref{eq:LLmodel}, which make it difficult to identify an appropriate microcanonical energy ``window'' encompassing many energy eigenstates while remaining narrow compared to the mean (postquench) energy $E$~[Eq.~\eqref{eq:E_final}].   We therefore consider the canonical ensemble (CE).  The density matrix of the CE is given by  
\begin{align}\label{eq:rho_CE}
    \hat{\rho}_\mathrm{CE} = Z_\mathrm{CE}^{-1}\sum_{\{\lambda_j\}} e^{-\beta E_{\{\lambda_j\}}}|\{\lambda_j\}\rangle\langle \{\lambda_j\}|,
\end{align}
where the inverse temperature $\beta$ is defined implicitly by $Z_\mathrm{CE}^{-1}\sum_{\{\lambda_j\}} \exp(-\beta E_{\{\lambda_j\}}) E_{\{\lambda_j\}} = E$ and the partition function $Z_\mathrm{CE} = \sum_{\{\lambda_j\}} \exp(-\beta E_{\{\lambda_j\}})$.  It is important to note that the only constraint (beyond that of fixed particle number) imposed in the CE is the conservation of the mean energy.  Thus, in contrast to the definition of $\hat{\rho}_\mathrm{DE}$ in Eq.~\eqref{eq:DEdmt}, the sum in Eq.~\eqref{eq:rho_CE} formally runs over all $N$-particle eigenstates $|\{\lambda_j\}\rangle$, regardless of parity and including those with nonzero values of the total momentum defined in Eq.~\eqref{eq:MomentumBR}~\cite{NoteH}.  Similarly to our calculations of DE expectation values, in practice we construct expectation values in the CE from a finite set of eigenstates, though we note that for a given level of accuracy their calculation requires us to include many more eigenstates than are required in the calculation of expectation values in the DE density matrix $\hat{\rho}_\mathrm{DE}$, as we discuss in Appendix~\ref{app:cutoff}.

\subsection{Momentum distribution}\label{subsec:DECEfirst}
\begin{figure}
    \includegraphics[width=0.48\textwidth]{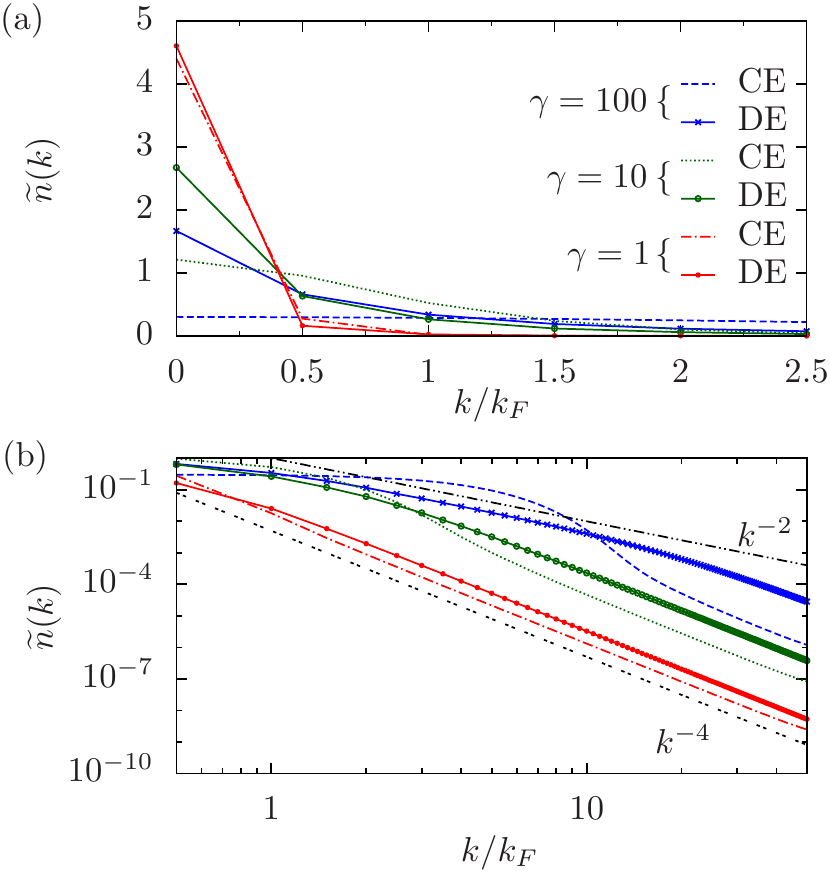}
    \caption{\label{fig:momdistGibbs} (Color online)  (a) Comparison of equilibrium momentum distributions $\widetilde{n}_\mathrm{DE}(k)$ and $\widetilde{n}_\mathrm{CE}(k)$ predicted by the DE and CE, respectively, for an interaction-strength quench of $N=5$ particles.  (b) The same momentum distributions on a double-logarithmic scale.  The black dotted line indicates the universal $\propto k^{-4}$ power-law scaling~\cite{Olshanii2003} observed at high momenta $k$.  For strong interactions, a power-law decay $\propto k^{-2}$ (black dot-dashed line) emerges at intermediate momenta.}
\end{figure}

In Fig.~\ref{fig:momdistGibbs}(a) we plot the DE momentum distribution $\widetilde{n}_\mathrm{DE}(k)$ for quenches of $N=5$ particles to final interaction strengths $\gamma=1,10,$ and $100$, along with the corresponding momentum distributions $\widetilde{n}_\mathrm{CE}(k)$ predicted by the CE.
Figure~\ref{fig:momdistGibbs}(b) shows the same momentum distributions on a logarithmic scale and reveals that for all interaction strengths, both $\widetilde{n}_\mathrm{DE}(k)$ and $\widetilde{n}_\mathrm{CE}(k)$ exhibit a power-law decay $\widetilde{n}(k)\propto k^{-4}$ (black dotted line) at high momenta~\cite{NoteI}.  This scaling behavior is a universal consequence of short-ranged two-body interactions in 1D~\cite{Olshanii2003,Caux2007,Barth2011} and indeed in higher dimensions~\cite{Tan2008,Braaten2012}. 

In the weakly excited case (Appendix~\ref{app:energy_and_finite_size_gap}) of a quench to $\gamma=1$, the DE (red solid line) and CE (red dot-dashed line) momentum distributions appear similar, with the zero-momentum occupation $\widetilde{n}_\mathrm{DE}(0)$ being only slightly larger than the corresponding CE value and the occupations $\widetilde{n}(k_{\pm 1})$ of the smallest magnitude nonzero momenta being somewhat smaller in the DE than in the CE.  From Fig.~\ref{fig:momdistGibbs}(b) we observe that in this case both the DE momentum distribution and that of the CE deviate from the $\propto k^{-4}$ power-law scaling (black dotted line) only at the smallest nonzero momenta resolvable in the finite periodic geometry.  In the relaxed (DE) state, our system is too small to observe the nontrivial long-wavelength behavior of the LL model for comparatively weak interactions $\gamma\lesssim 1$.  In fact, many low-lying excitations of the LL system that would be excited by a quench to $\gamma=1$ in an infinite system are not present in our finite-sized system.  As a result our system is only weakly excited above the ground state of $\hat{H}$ by the quench and the relaxation dynamics associated with the dephasing of energy eigenstates are not observed.  This results, in particular, in the near-monochromatic oscillations of $\widetilde{n}(0,t)$ for this quench, as discussed in Sec.~\ref{subsec:Fidelity} and Appendix~\ref{app:energy_and_finite_size_gap}.  

We note from Fig.~\ref{fig:momdistGibbs}(a) that the zero-momentum occupation $\widetilde{n}_\mathrm{DE}(0)$ in the DE and the prediction $\widetilde{n}_\mathrm{CE}(0)$ of the CE for this quantity both decrease significantly with increasing final interaction strength $\gamma$.  However, the decrease in $\widetilde{n}_\mathrm{CE}(0)$ with increasing $\gamma$ is much more pronounced than the corresponding decrease in $\widetilde{n}_\mathrm{DE}(0)$, and $\widetilde{n}_\mathrm{DE}(0)$ therefore exceeds $\widetilde{n}_\mathrm{CE}(0)$ by an increasingly large margin as $\gamma$ increases.  Figure~\ref{fig:momdistGibbs}(a) also reveals conspicuous differences, at larger values of $\gamma$, between the width and the shape of $\widetilde{n}_\mathrm{DE}(k)$ and those of $\widetilde{n}_\mathrm{CE}(k)$.  In particular, $\widetilde{n}_\mathrm{DE}(k)$ remains convex on $k\geq 0$ for all considered final interaction strengths, whereas $\widetilde{n}_\mathrm{CE}(k)$ develops an increasingly broad concave hump at small $k$ (cf. Ref.~\cite{Vignolo2013}) with increasing $\gamma$.  For $\gamma=100$ the width (half width at half maximum) of the CE momentum distribution is much greater than $k_F$, whereas $\widetilde{n}_\mathrm{DE}(k)$ is comparatively sharply peaked around $k=0$.  We observe from Fig.~\ref{fig:momdistGibbs}(b) that a scaling $\propto k^{-2}$ (black dot-dashed line) emerges at intermediate momenta for $\gamma \sim 100$.  This same power-law scaling has been obtained analytically~\cite{Kormos2014} in the singular limit of a quench to the TG limit of infinitely strong interactions, where it was found to persist in the limit $k\to\infty$.  By contrast, the universal $\propto k^{-4}$ scaling of the momentum distribution at large $k$~\cite{Olshanii2003,Tan2008,Barth2011} is always observed in the quenches to finite final interaction strengths $\gamma$ that we consider here.      

We remark that at comparatively low temperatures, such that the LL system is in the quantum-degenerate regime, the known asymptotic form of the thermal-equilibrium first-order correlation function $g^{(1)}(x)$ at large separations $x$ is an exponential decay~\cite{Bogoliubov2004,Cazalilla2004,Cazalilla2011}, corresponding to a Lorentzian functional form for $\widetilde{n}(k)$ at small $k$.  At increasingly higher temperatures, the effects of both interactions and particle statistics eventually become negligible, and $g^{(1)}(x)$ becomes Gaussian with width given by the thermal de~Broglie wavelength (see, e.g., Ref.~\cite{Lenard1966}), corresponding to a Gaussian momentum distribution $\widetilde{n}(k)$ that becomes increasingly broad with increasing temperature.  Although Fig.~\ref{fig:momdistGibbs} indicates that $\widetilde{n}_\mathrm{CE}(k)$ is consistent with these known thermal-equilibrium results, the momentum distributions $\widetilde{n}_\mathrm{DE}(k)$ we observe here show a qualitatively distinct behavior.  In particular, for $\gamma=100$, the Gaussian form of $\widetilde{n}_\mathrm{CE}(k)$ demonstrates that the energy imparted to the system by the quench, if redistributed during relaxation so as to agree with the principles of conventional statistical mechanics, would heat the system to temperatures far above quantum degeneracy.  By contrast, the DE momentum distribution $\widetilde{n}_\mathrm{DE}(k)$ appears to retain the Lorentzian-like character expected for the LL model at nonzero but small temperatures, such that quantum-degeneracy effects remain significant.  We note also that the coefficient $\lim_{k\to\infty} k^4\widetilde{n}(k)$ of the high-momentum tail (i.e., the Tan contact~\cite{Olshanii2003,Tan2008,Barth2011}) in the DE is always larger than that in the CE.  In the case of $\gamma=1$ this coefficient is larger in the DE as compared to the CE by a factor of approximately two, and its value in the DE exceeds that in the CE by an increasingly large factor as $\gamma$ increases, being more than an order of magnitude larger in the case of $\gamma=100$.

\subsection{Second-order correlations}
\begin{figure}
    \includegraphics[width=0.48\textwidth]{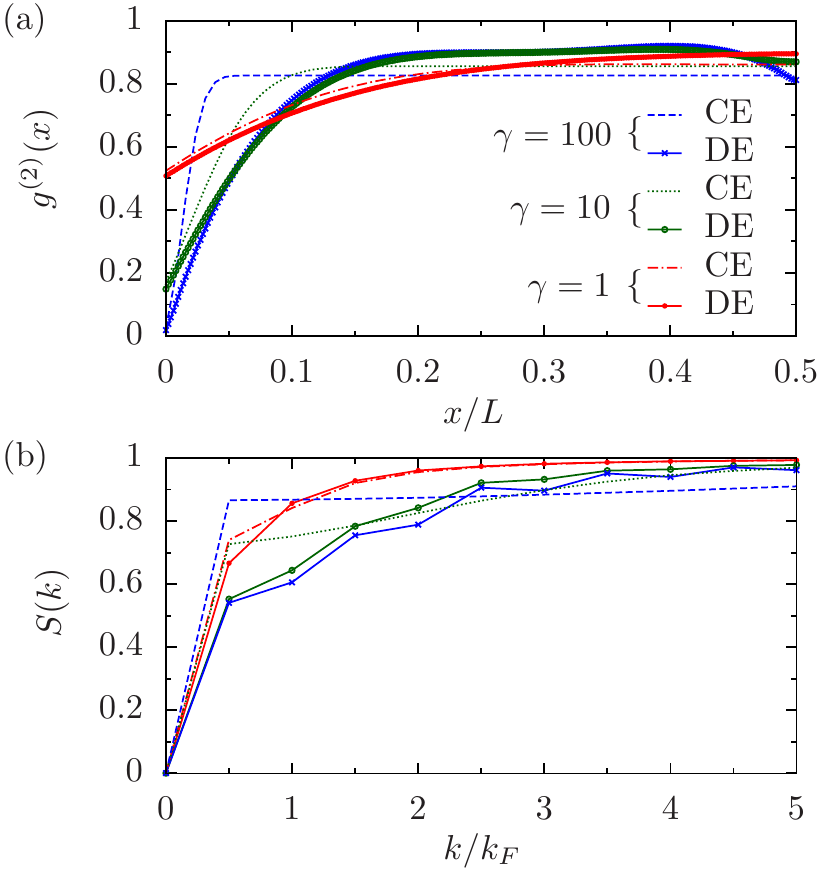}
    \caption{\label{fig:g2DE} (Color online)  Second-order correlations in the DE for quenches of $N=5$ particles to $\gamma = 1, 10,$~and~$100$.  (a) Second-order correlation function $g^{(2)}_\mathrm{DE}(x)$ and (b) corresponding structure factor $S_\mathrm{DE}(k_j)$.  The legend is the same for both panels and is indicated in (a).}
\end{figure}

In Fig.~\ref{fig:g2DE}(a) we plot the predictions $g^{(2)}_\mathrm{DE}(x)$ of the DE for the equilibrium second-order correlations of the postquench system, along with the corresponding predictions $g^{(2)}_\mathrm{CE}(x)$ of the CE for this quantity.  For $\gamma=1$ the nonlocal real-space correlation function $g^{(2)}_\mathrm{DE}(x)$ [small red circles in Fig.~\ref{fig:g2DE}(a)] is similar to the CE form $g^{(2)}_\mathrm{CE}(x)$ (red dot-dashed line), and both are comparable to the form of $g^{(2)}(x)$ found for $\gamma\lesssim 1$ at zero temperature in previous works~\cite{Astrakharchik2003,Astrakharchik2006,Cherny2009,Sykes2008}, consistent with the weak excitation of the system observed in the behavior of the momentum distribution (Sec.~\ref{subsec:DECEfirst}) for this final interaction strength.  We note that both the local second-order coherence $g^{(2)}_\mathrm{DE}(0)$ in the DE and that in the CE decrease significantly as $\gamma$ is increased.  However, the ``Friedel'' oscillations of wavelength $\sim 1/k_F$ that appear in $g^{(2)}(x)$ for strong interaction strengths $\gamma \gg 1$ at zero temperature~\cite{Korepin1993,Cherny2006,Sykes2008} are not seen in either the DE or the CE predictions for the equilibrium second-order coherence at large values of $\gamma$.  Indeed for $\gamma=10$~and~$100$ the results for $g^{(2)}_\mathrm{DE}(x)$ are qualitatively similar to the behavior of the second-order coherence in the high-temperature fermionization regime~\cite{Kheruntsyan2003,Gangardt2003b,Sykes2008}, consistent with the results of the lattice-model simulations of Ref.~\cite{Muth2010a} and studies of quenches to the TG limit~\cite{Gritsev2010,Kormos2013,Kormos2014,DeNardis2014a}.  We note, however, that the dip in the second-order correlation function about $x=0$ is significantly wider in the DE than in the CE for $\gamma=10$~and~$100$.  Moreover, for these large final interaction strengths the function $g^{(2)}_\mathrm{DE}(x)$ is not completely flat outside the central ``fermionic'' dip at small $x$ and, in fact, as the separation $x$ approaches the midpoint $L/2$ of the periodic geometry, the second-order coherence exhibits a small secondary dip to a value lower than the roughly constant value of $g^{(2)}_\mathrm{DE}(x)$ at intermediate separations.  We have found that this feature is highly sensitive to the particle number $N$, varying between a small dip (as seen here) and a small peak for odd and even values of $N$, respectively, and we therefore identify it as a finite-size artifact that should gradually vanish with increasing system size.

Figure~\ref{fig:g2DE}(b) shows the DE predictions for the equilibrium structure factors $S_\mathrm{DE}(k)$ obtained from the correlation functions $g^{(2)}_\mathrm{DE}(x)$ plotted in Fig.~\ref{fig:g2DE}(a) via Eq.~\eqref{eq:structfact}, along with the corresponding CE structure factors $S_\mathrm{CE}(k)$.  Unsurprisingly, for $\gamma=1$ this representation of the second-order correlations in the DE is also similar to the predictions of the CE, whereas for both of the larger values of $\gamma$ we consider, the DE prediction $S_\mathrm{DE}(k)$ differs markedly from $S_\mathrm{CE}(k)$ and also from the corresponding zero-temperature form of the structure factor (see, e.g., Refs.~\cite{Caux2006,Cherny2009}).  In particular, the DE predictions for these structure factors have smaller magnitudes at small momenta $k\lesssim k_F$ than the corresponding CE structure factors.  We note that our results for the equilibrium static structure factor following the quench are at least qualitatively similar to those of Refs.~\cite{DeNardis2014a,Kormos2014}, aside from the obvious distinction that the characteristic $\gamma$-independent value $S(0)=1/2$ obtained in Ref.~\cite{DeNardis2014a} is precluded in our calculations by particle-number conservation, which imposes $S_\mathrm{DE}(0)=0$~\cite{NoteJ}.

\subsection{Local correlations}
\begin{figure}
    \includegraphics[width=0.48\textwidth]{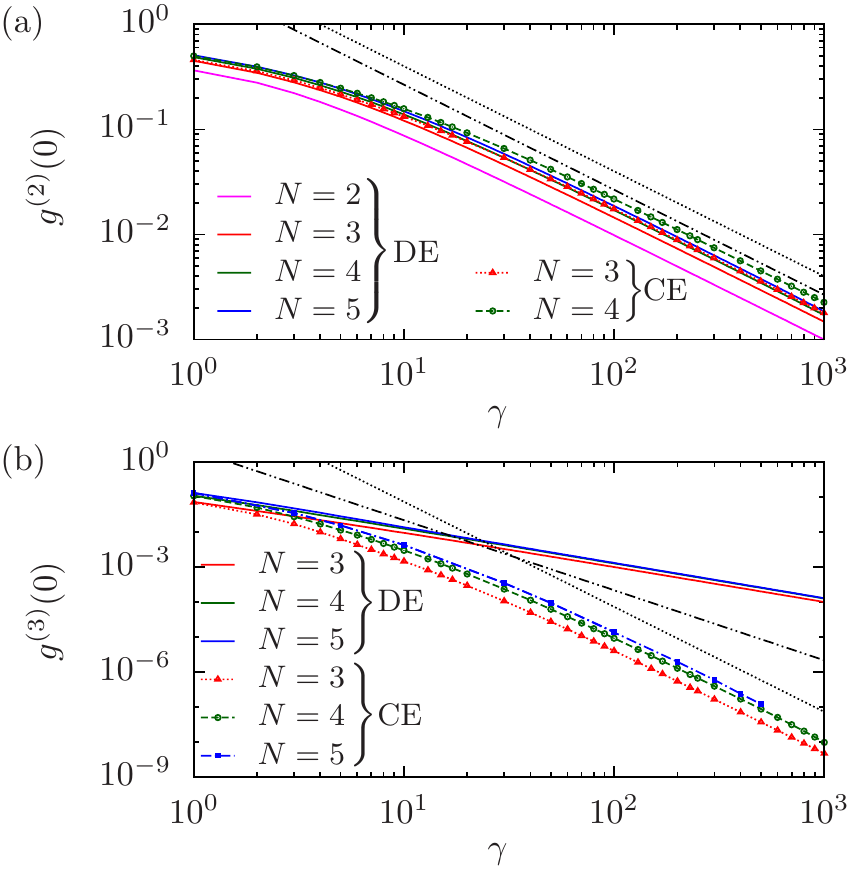}
    \caption{\label{fig:g23localvsgamma} (Color online)  Comparison of equilibrium values of local correlation functions predicted by the DE and the CE.  (a) Local second-order coherence functions $g^{(2)}_\mathrm{DE}(0)$ and $g^{(2)}_\mathrm{CE}(0)$.  (b) Local third-order coherence functions $g^{(3)}_\mathrm{DE}(0)$ and $g^{(3)}_\mathrm{CE}(0)$.  In both panels, black dotted and dot-dashed lines indicate thermodynamic-limit predictions for the corresponding correlation functions obtained in the grand-canonical ensemble and the generalized TBA calculations of Refs.~\cite{Kormos2013,DeNardis2014a}, respectively (see text).}
\end{figure}

We now compare the DE values $g^{(2)}_\mathrm{DE}(0)$ and $g^{(3)}_\mathrm{DE}(0)$ of the local second- and third-order correlation functions, respectively, to the predictions of the CE for these quantities.  The dramatically reduced computational expense involved in calculating local correlation functions, as compared to nonlocal correlation functions such as $g^{(1)}(x)$ and $g^{(2)}(x)$, allows us to pursue our investigations to much larger values of $\gamma$ than we have considered so far while maintaining a comparable level of accuracy (see Appendix~\ref{app:cutoff}).  We therefore present in Fig.~\ref{fig:g23localvsgamma} results for $g^{(2)}_\mathrm{DE}(0)$ and $g^{(3)}_\mathrm{DE}(0)$ for final interaction strengths up to $\gamma=10^3$.   

In Fig.~\ref{fig:g23localvsgamma}(a) we plot $g^{(2)}_\mathrm{DE}(0)$ for $N=2,3,4,$ and $5$ particles (solid lines, bottom to top), together with the thermal-equilibrium values $g^{(2)}_\mathrm{CE}(0)$ obtained in the canonical ensemble, for $N=3$ and $4$ particles (red triangles and green circles, respectively).  We observe that both ensembles predict $g^{(2)}(0)$ to exhibit behavior consistent with power-law decay $\propto 1/\gamma$ at large values of $\gamma$, though for any given value of $\gamma$ and particle number $N$, the DE result $g^{(2)}_\mathrm{DE}(0)$ is somewhat smaller than $g^{(2)}_\mathrm{CE}(0)$.  This behavior is consistent with the results of the generalized TBA calculations of Refs.~\cite{Kormos2013,DeNardis2014a}, which both predict an asymptotic form $g^{(2)}_\mathrm{GTBA}(0)\sim 8/(3\gamma)$ (black dot-dashed line) for the local second-order coherence following a quench of the LL-model interaction strength from zero to $\gamma$.  As noted in Ref.~\cite{Kormos2013}, this prediction for the equilibrium postquench value of $g^{(2)}(0)$ has the same power-law scaling exponent as the corresponding prediction $g^{(2)}_\mathrm{GCE}(0)\sim 4/\gamma$ of the grand-canonical ensemble~\cite{Kormos2013,Gangardt2003b,NoteK} (black dotted line), but a significantly smaller prefactor.
We note not only that $g^{(2)}_\mathrm{DE}(0)$ here exhibits the same $\propto 1/\gamma$ scaling as $g^{(2)}_\mathrm{CE}(0)$ and that its prefactor is indeed smaller, but also that our results for $g^{(2)}_\mathrm{DE}(0)$ and $g^{(2)}_\mathrm{CE}(0)$ appear to be scaling towards the asymptotic predictions of Ref.~\cite{Kormos2013,DeNardis2014a} for $g^{(2)}(0)$ in the generalized statistical ensembles considered in those works and the grand-canonical ensemble, respectively, as the particle number $N$ is increased. 

We now turn our attention to the local third-order correlation functions $g^{(3)}_\mathrm{DE}(0)$ and $g^{(3)}_\mathrm{CE}(0)$, which we plot in Fig.~\ref{fig:g23localvsgamma}(b) for $N=3,4,$ and $5$ particles (solid lines and symbols, respectively).  We observe that for all three particle numbers, the behavior of $g^{(3)}_\mathrm{DE}(0)$ is consistent with power-law scaling $\propto \gamma^{-1}$ at large $\gamma$, in pronounced disagreement with the prediction $g^{(3)}_\mathrm{GTBA}(0)\sim32/(15\gamma^2)$ of Refs.~\cite{Kormos2013,DeNardis2014a} (black dot-dashed line).  By contrast, the results of our CE calculations appear to be scaling towards the grand-canonical prediction~\cite{NoteK} $g^{(3)}_\mathrm{GCE}(0)\sim 72/\gamma^3$ (black dotted line) with increasing $N$. 

Although we employ a sufficiently large basis of LL eigenstates in our calculation of DE expectation values that the values of the local coherences appear reasonably insensitive to the precise number of states we use, the accuracy of our results for the local coherences is inevitably limited by this eigenstate ``cutoff'' (see Appendix~\ref{app:cutoff}).  However, we stress that the local correlation function $g^{(3)}(0)$ is [like $g^{(2)}(0)$] non-negative in any LL eigenstate $|\{\lambda_j\}\rangle$, and raising the cutoff to include some or all of the weakly occupied eigenstates omitted in our numerical calculation of this quantity could therefore only \emph{increase} its value.  Moreover, the total occupation of neglected eigenstates in our DE calculations increases with increasing $\gamma$ (Appendix~\ref{app:cutoff}).  Thus, we expect our calculated value of $g^{(3)}_\mathrm{DE}(0)$ to increasingly \emph{underestimate} the exact value of this quantity with increasing $\gamma$; i.e., the scaling $g^{(3)}_\mathrm{DE}(0)\propto \gamma^{-1}$ shown in Fig.~\ref{fig:g23localvsgamma}(b) should constitute an \emph{upper} bound to the rate at which $g^{(3)}_\mathrm{DE}(0)$ scales to zero, whereas the prediction of Refs.~\cite{Kormos2013,DeNardis2014a} vanishes even more rapidly.  Of course, our results here are for strongly finite-sized systems of at most $N=5$ particles, and the reader might expect that the discrepancy between $g^{(3)}_\mathrm{DE}(0)$ and the results of Refs.~\cite{Kormos2013,Kormos2014} should disappear in the thermodynamic limit.  However, results for local correlation functions at zero temperature~\cite{Cheianov2006,Zill2014} and our results for $g^{(2)}_\mathrm{DE}(0)$ [Fig.~\ref{fig:g23localvsgamma}(a)] both suggest that local correlations, and in particular their scaling with interaction strength, become increasingly insensitive to finite-size effects as the TG limit is approached.  We note that the power-law behavior $g^{(3)}_\mathrm{GTBA}(0) \propto \gamma^{-2}$ obtained in the calculations of Ref.~\cite{Kormos2013,DeNardis2014a} lies in between the thermal scaling $g^{(3)}_\mathrm{GCE}(0)\propto\gamma^{-3}$ and the result $g^{(3)}_\mathrm{DE}(0)\propto\gamma^{-1}$ of our DE calculations.  We remark that this may be an indication that the GGE and quench-action calculations of Refs.~\cite{Kormos2013}~and~\cite{DeNardis2014a}, respectively, only partially account for the constraints to which the integrable LL system is subject.  The origin of this discrepancy remains an important question for future study.

\section{Summary}\label{sec:summary}
We have investigated the dynamics of the Lieb--Liniger model of 1D contact-interacting bosons following a sudden quench of the interaction strength from zero to a positive value.  We computed the long-time evolution of systems containing up to five particles by expanding the time-evolving pure-state wave function of the postquench system over a truncated basis consisting of all energy eigenstates with (absolute) overlap with the initial state of the system larger than a chosen threshold.  These overlaps, and the matrix elements of observables between energy eigenstates, were obtained by symbolic evaluation of the corresponding coordinate-space integrals in terms of the rapidities that label the states, which were themselves obtained as numerical solutions of the appropriate Bethe equations.  

We found that for quenches to comparatively small final interaction strengths ($\gamma \lesssim 1$), observables exhibit near-monochromatic oscillations.  We identified this as a consequence of the gap in the energy spectrum induced by the finite size of the system, which severely suppresses the excitation of the system for small values of the final interaction strength, resulting in quasi-two-level system dynamics.  For stronger interaction strengths, we observed results for the first- and second-order correlations consistent with the relaxation of the integrable many-body system due to the dephasing of the $N$-particle energy eigenstates.  We also observed the propagation of correlation waves in the second-order correlations of the system, which are related to density modulations.  We found that the behavior of the fidelity between the initial (prequench) state and the state at time $t$ following the quench is qualitatively similar to that of nonlocal quantities such as the occupation of the zero-momentum single-particle mode, indicating that these experimentally relevant quantities provide effective probes of the eigenstate dephasing of the $N$-body system.  Local correlations, however, decay much more rapidly and do not necessarily reflect the relaxation of the system as a whole.    

We assessed the character of correlations in the relaxed state by comparing diagonal-ensemble correlations to those of the canonical ensemble, in which only the conservation of energy and normalization are taken into account.  In particular, we observed that for quenches to large $\gamma$, the relaxed state of the system exhibits a momentum distribution consistent with the asymptotically Lorentzian form expected for the Lieb--Liniger model at low-temperature thermal equilibrium.  This is in stark contrast to the canonical-ensemble prediction for the relaxed postquench state, which yields a Gaussian momentum distribution consistent with temperatures well above quantum degeneracy.  Our calculations also indicate that in the Tonks--Girardeau limit $\gamma\to\infty$ the local second-order coherence $g^{(2)}_\mathrm{DE}(0)$ scales towards zero with the same power law as the corresponding correlation function in the canonical ensemble (i.e., like $1 / \gamma$), but with a smaller prefactor, consistent with the results of Refs.~\cite{Kormos2013,DeNardis2014a}.  However, although our results for the local third-order coherence in the canonical ensemble are consistent with the expected behavior of a thermal system, our results for $g^{(3)}(0)$ in the nonthermal diagonal ensemble show a scaling $\propto \gamma^{-1}$, slower than both the $\propto \gamma^{-3}$ scaling expected for a thermal state and the $\propto \gamma^{-2}$ scaling predicted by the generalized thermodynamic Bethe-ansatz calculations of Refs.~\cite{Kormos2013,DeNardis2014a}.  Whether this discrepancy is merely a consequence of the finite size of our system or is indicative of subtleties not captured in the methodologies of Refs.~\cite{Kormos2013,DeNardis2014a} is an important question for further study.

\begin{acknowledgments}
We wish to acknowledge discussions with H. Buljan, \mbox{J.-S.} Caux, F.~H.~L. Essler, and M. Rigol.  J.C.Z. would like to thank E. Bittner, K. Schade, and C. Feng for technical help with computational tasks.  This work was supported by Australian Research Council Discovery Project No. DP110101047 (J.C.Z., T.M.W., K.V.K., and M.J.D.), by the Deutsche Forschungsgemeinschaft, Grant No. GA677/7,8 (T.G.), the University of Heidelberg (Center for Quantum Dynamics), and the Helmholtz Association (Grant No. HA216/EMMI) (J.C.Z. and T.G.).
\end{acknowledgments}

\appendix

\section{Basis-set truncation}\label{app:cutoff}
Expression~\eqref{eq:initialspectral} for $|\psi(t)\rangle$ [and, consequently, Eqs.~\eqref{eq:observabletime}--\eqref{eq:DEdmt} derived from it] involves a sum $\sum_{\{\lambda_j\}}$ over all zero-momentum, parity-invariant states $|\{\lambda_j\}\rangle$.  In principle, there are an infinite number of such states that contribute to the sum.  However, in practical numerical calculations, we must truncate the sum to a finite number of terms in some manner.  The accuracy of our calculations based on this truncated sum can then be quantified by the sum rules satisfied by the conserved quantities of the system.  We focus primarily on the normalization sum rule $\sum_{\{\lambda_j\}} |C_{\{\lambda_j\}}|^2 = 1$ (cf. Ref.~\cite{Mossel2010}).  

\begin{table}[t]
    \centering
    \caption{\label{tab:cutoffs1}  Basis-set sizes and sum-rule violations for time-evolving correlations and statistical-ensemble expectation values.  Energy cutoff $E_\mathrm{cut}$ applies only for CE calculations, and the CE density matrix defined in Eq.~\eqref{eq:rho_CE} automatically satisfies the normalization sum rule.}
    \begin{tabular}{cccccc}
    \hline \hline
    $\gamma $ & $\;$ Type$\,^\mathrm{a}$ $\;$ & No. states & $\quad \Delta N \quad$ & $\quad\enskip \Delta E \quad\enskip$ & $\enskip E_\mathrm{cut}/k_F^2 \enskip$ \\
    \hline 
    $1$ & $\langle \hat{O}(t) \rangle$ & $1221$ & $5\times 10^{-8}$ & $2\times 10^{-3}$ & N/A \\
    $1$ & DE & $6770$ & $4\times 10^{-10}$ & $5\times 10^{-4}$ & N/A \\
    $1$ & CE & $3.7\times10^6$ & N/A & $2\times 10^{-7}$ & $4.0\times10^2$ \\

    $10$ & $\langle \hat{O}(t) \rangle$ & $1221$ & $7\times 10^{-6}$ & $2\times 10^{-2}$ & N/A \\
    $10$ & DE & $6770$ & $8\times 10^{-8}$ & $5\times 10^{-3}$ & N/A \\
    $10$ & CE & $3.7\times10^6$ & N/A & $8\times 10^{-6}$ & $4.0\times10^2$ \\
    
    $100$ & $\langle \hat{O}(t) \rangle$ & $1221$ & $10^{-3}$ & $2\times 10^{-1}$ & N/A \\
    $100$ & DE & $6770$ & $3\times 10^{-5}$ & $5\times 10^{-2}$ & N/A \\
    $100$ & CE & $3.7\times10^6$ & N/A & $8\times 10^{-6}$ & $4.0\times10^2$ \\

    \hline\hline

    \end{tabular}
    \flushleft
    $^\mathrm{a}$ Fidelities $F(t)$ are calculated from the DE basis sets.
\end{table}

In our calculations we include all states $|\{\lambda_j\}\rangle$ for which the absolute overlap $|\langle\{\lambda_j\}|\psi_0\rangle|$ with the initial state [Eq.~\eqref{eq:psiini}] is larger than some threshold value.  Our approach exploits the fact that the solutions $\{\lambda_j\}$ of the Bethe equations~\eqref{eq:Betheeq} are in one-to-one correspondence~\cite{Yang1969} with the (half-)integers $\{m_j\}$ that appear in Eq.~\eqref{eq:Betheeq}.  As the states $|\{\lambda_j\}\rangle$ are parity invariant, we can choose to label the rapidities such that $\lambda_j=-\lambda_{N+1-j}$, where $\lambda_1>\lambda_2>\cdots>\lambda_N$.  Then we can label the states simply by $(m_1,m_2,\dots,m_{\lfloor (N+1)/2\rfloor})$, where $\lfloor x \rfloor$ denotes the integer part of $x$.  We specialize hereafter to the case $N=5$, which is the largest $N$ for which we consider the dynamics in this article.  Our approach reduces in a natural way to the cases of $N\leq 4$.  The states can be grouped into families labeled by $m_1=2,3,\dots$, where within each family the second quantum number can assume values $1\leq m_2 < m_1$ (and $m_3=0$).  We have found from our explicit evaluation of the overlaps~\cite{Zill2014} that $|\langle \{\lambda_j\} | \psi_0\rangle|$ decreases monotonically with increasing $m_2$ within each family $m_1$ and, moreover, that the first member $(m_1,1,0)$ of each family $m_1$ has a larger (absolute) overlap with $|\psi_0\rangle$ than the first member $(m_1+1,1,0)$ of the next family~\cite{DeNardis2014a,Brockmann2014a,*Brockmann2014b,*Brockmann2014c,Zill2014,NoteL}.  We therefore construct the basis by considering in turn each family $m_1$ and including all states within that family for which the overlap with the initial state exceeds our chosen threshold value.  Eventually, for some value of $m_1$, even the first state $(m_1,1,0)$ of the family has overlap with $|\psi_0\rangle$ smaller than the threshold, at which point all states that meet the overlap threshold have been exhausted.  The basis so constructed therefore comprises the $\mathcal{N}$ states with the largest overlap with $|\psi_0\rangle$ and thus minimizes the violation $\Delta N = 1-\sum_{\{\lambda_j\}} |C_{\{\lambda_j\}}|^2$ of the normalization sum rule for this basis size. 

\begin{table}[b]
    \centering
    \caption{\label{tab:cutoffs2}  Basis-set sizes and sum-rule violations for the local correlation functions plotted in Fig.~\ref{fig:g23localvsgamma}.}
    \begin{tabular}{cccccc}
    \hline \hline
    $N$ & $\;$ Type $\;$ & No. states & $\,\,\quad \Delta N\,^\mathrm{a} \!\!\quad$ & $\,\quad\enskip \Delta E\,^\mathrm{a} \!\quad\enskip$ & $\enskip E_\mathrm{cut}/k_F^2 \enskip$ \\
    \hline 
    $3$ & DE & $10^4$ & $10^{-8}$ & $2\times 10^{-2}$ & N/A \\
    $3$ & CE & $3.9\times10^5$ & N/A & $10^{-6}$ & $4.8\times10^3\,^\mathrm{\phantom{b}}$ \\

    $4$ & DE & $9.5\times10^4$ & $3\times 10^{-6}$ & $5\times 10^{-3}$ & N/A \\ 
    $4$ & CE & $3.2\times10^6$ & N/A & $10^{-6}$ & $1.6\times 10^3\,^\mathrm{\phantom{b}}$ \\
    
    $5$ & DE & $1.9\times10^5$ & $5\times 10^{-6}$ & $5\times 10^{-2}$ & N/A \\
    $5$ & CE & $5.9\times10^6$ & N/A & $5\times 10^{-7}$ & $4.8\times10^2\,^\mathrm{b}$ \\

    \hline\hline
    \end{tabular}
    \flushleft
    $^\mathrm{a}$ Sum-rule discrepancies quoted are those for $\gamma=10^3$ ($\gamma=5\times10^2$ for $\Delta E$ in the $N=5$ CE). \\
    $^\mathrm{b}$ For quenches to $\gamma < 50$, cutoff energy $E_\mathrm{cut}=4.0\times10^2\,k_F^2$.
\end{table}

For an integrable system such as we consider here, the normalization is just one of an infinite number of sum rules defined by the conserved quantities $Q^{(m)} = \sum_j (\lambda_j)^m$ of the LL Hamiltonian~\eqref{eq:LLmodel}.  However, all the odd charges $Q^{(2n+1)}$, with $n$ an integer, are zero by the constraint to parity-invariant states.  Moreover, even charges $Q^{(2n)}$ with $2n\geq 4$ are formally singular~\cite{Kormos2013}, diverging as any rapidity $\lambda_k \in \{\lambda_j\}$ is increased toward infinity.  Thus, the only nontrivial and regular conserved quantity other than the normalization is the energy $\langle\hat{H}\rangle = \sum_{\{\lambda_j\}} |C_{\{\lambda_j\}}|^2 \sum_k (\lambda_k)^2$ [cf. Eq.~\eqref{eq:EnergyBR}].  We note that this quantity converges as $1/ \lambda_j$, which is much slower than the $\propto 1/ \lambda_j^3$ convergence of the normalization.  We characterize the saturation of this sum rule by the energy sum-rule violation $\Delta E = (E - \sum_{\{\lambda_j\}} |C_{\{\lambda_j\}}|^2 \sum_k (\lambda_k)^2)/E$, where $E$ is the exact postquench energy [Eq.~\eqref{eq:E_final}].  As a consequence of the slow convergence of the energy with increasing basis-set size, the energy sum rule is, in general, less well satisfied in our calculations than the normalization sum rule. 

\begin{figure}[t]
    \includegraphics[width=0.48\textwidth]{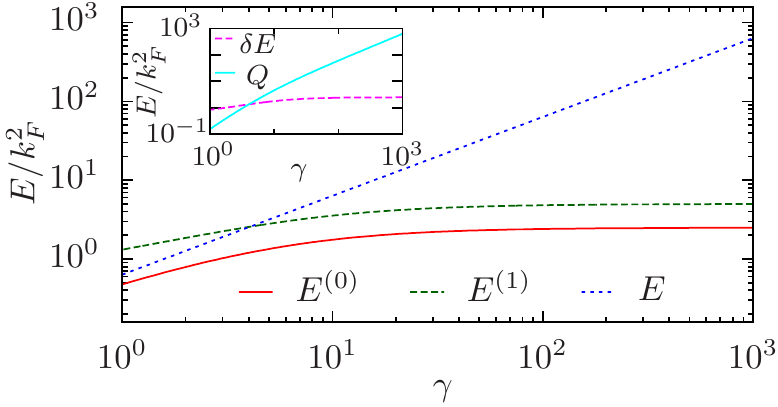}
    \caption{\label{fig:energies} (Color online) Energy of a system of $N=5$ particles following a quench of the interaction strength from zero to $\gamma>0$.  For comparison, the energy $E^{(0)}$ of the ground state of $\hat{H}$ and the energy $E^{(1)}$ of the lowest-lying excited state of $\hat{H}$ that has finite overlap with the initial state $|\psi_0\rangle$ are also shown.  Inset: Heat $Q$ added to the system by the quench and the energy gap $\delta E$ between the ground state and the lowest-lying state that has finite overlap with the initial state (see text).}
\end{figure}

We note also that the evaluation of time-dependent observables [Eq.~\eqref{eq:observabletime}] involves a double summation over $\{\lambda_j\}$ and is thus more numerically demanding than the calculation of correlations in the DE [Eq.~\eqref{eq:DE}], for which only a single sum occurs (i.e., only diagonal elements contribute).  An exception is the time-evolving fidelity $F(t)$, which can be written as the modulus square of a single sum over eigenstates [cf. Eq.~\eqref{eq:fidelity}].   We list the sizes of the basis sets employed in our calculations, together with the resulting violations $\Delta N$ and $\Delta E$ of the norm and energy sum rules, respectively, in Table~\ref{tab:cutoffs1}.  For expectation values in the CE [Eq.~\eqref{eq:rho_CE}], the truncation of the basis set is most appropriately performed by retaining all states with energy $E$ below some cutoff energy $E_\mathrm{cut}$.  The inverse temperature $\beta$ is then chosen as that which, within a prescribed tolerance level, minimizes the energy sum-rule violation $\Delta E$.  In this case, the sum is not restricted to parity-invariant, or even zero-momentum, states.  However, the weights of states in the ensemble decrease exponentially with energy, and we have found that the energy cutoffs used in our CE calculations, which we also list in Table~\ref{tab:cutoffs1}, are sufficiently large to ensure saturation of the momentum distributions plotted in Fig.~\ref{fig:momdistGibbs}.   

The results for the local second- and third-order correlation functions presented in Fig.~\ref{fig:g23localvsgamma} constitute a more demanding test of numerical accuracy, due to the large values of $\gamma$ considered.  We list the sizes of the basis sets used in these calculations and the resulting sum-rule violations in Table~\ref{tab:cutoffs2}.

\section{Post-quench energy and finite-size gap}\label{app:energy_and_finite_size_gap} 
In Fig.~\ref{fig:energies} we plot the postquench energy $E$ [Eq.~\eqref{eq:E_final}] as a function of the final interaction strength $\gamma$ (blue dotted line).  
For comparison, we also plot the energy $E^{(0)}(\gamma)$ of the ($N$-particle) ground state of the LL Hamiltonian~\eqref{eq:LLmodel} with interaction strength $\gamma$ (red solid line).  The difference between these two energies, $Q\equiv E - E^{(0)}(\gamma)$, can be identified as the heat added to the system by the quench~\cite{Polkovnikov2008}, which we plot in the inset to Fig.~\ref{fig:energies} (cyan solid line).  

We note that although the excitation spectrum of the LL system is gapless in the thermodynamic limit, in a finite-sized system a gap of order $1/L$~\cite{Korepin1993} (and thus $\sim 1/N$ at fixed density) between the energies of the ground state and the lowest-lying excited state(s) appears.  In Fig.~\ref{fig:energies} we plot (green dashed line) the energy $E^{(1)}$ of the lowest-lying state that has finite overlap with the initial state (see Sec.~\ref{sec:dynamics}).  We observe that the gap $\delta E=E^{(1)}-E^{(0)}$ between this energy and that of the ground state is $\sim 2 k_F^2$ for the system sizes we consider (magenta dashed line in inset to Fig.~\ref{fig:energies}).   We note that for large $\gamma\gg 1$, the heat $Q$ added to the system is much larger than the finite-size gap $\delta E$, whereas for $\gamma \lesssim 10$ the two energies are comparable, and for $\gamma \sim 1$, the gap is, in fact, larger than the added heat $Q$.  It is clear, therefore, that in this regime the system can only be weakly excited above the ground state of $\hat{H}$ by the quench, due to the presence of the finite-size gap.  Thus, in quenches to $\gamma=1$, we observe almost purely monochromatic oscillations of observables, as many low-lying excitations of the formally gapless system are not present in the finite geometry and the dynamics of the system are dominated by the two most highly occupied eigenstates of $\hat{H}$.  By contrast, for large values of $\gamma\gtrsim 10$, the finite-size gap is relatively small compared to the energy imparted to the system during the quench, and as a result many energy eigenstates contribute significantly to the postquench dynamics.  Thus, for quenches to large values of $\gamma$ many states are available to realize the eigenstate-dephasing picture of relaxation dynamics, consistent with the results of our calculations~\cite{NoteE}.

\bibliographystyle{prsty}

\begin{thebibliography}{100}
\bibitem{NoteD}
{We quote the speed of sound here in units of the Fermi wave vector $k_F$ of
  our finite-sized system (Sec.~\ref{subsec:correlation_functions}), which
  differs from the Fermi wave vector $k_F^{(\infty)} = n\pi$ in the
  thermodynamic limit by an $O(1/N)$ correction. We note that this finite-size
  correction is here much larger than the $O(1/\gamma)$ strong-coupling
  correction to the TG-limit speed of sound.}
\vspace{0.1\baselineskip}
\bibitem{NoteG}
{We note that the ratio of the minimum velocity $v_\mathrm{min}$ to the Fermi
  velocity $v_F$ depends, in general, on the particle number $N$. We have found
  that for $N=4$, density dips propagate at velocities consistent with integer
  multiples of $v_\mathrm{min}$, which correspond to integer multiples of
  $v_F/2$.}
\vspace{0.1\baselineskip}
\bibitem{NoteH}
{Although one could consider more refined definitions of the CE that are
  restricted so as to involve only states that have finite overlap with the
  initial state---e.g., states with zero total momentum or strictly
  parity-invariant states---we have found that these refined CE definitions do
  not yield correlation functions that agree more closely with the DE results.
  Therefore, for clarity, we simply take as our ``reference'' thermal ensemble
  the most common definition of CE, in which, at fixed particle number $N$, all
  conservation laws other than conservation of energy are ignored.}
\vspace{0.1\baselineskip}
\bibitem{NoteI}
{We note that the accuracy with which we can resolve correlations at high
  values of $k$ is ultimately limited by the density of the (Cartesian)
  position-space grid on which we calculate correlation functions. We stress
  that this effective ``momentum cutoff'' is independent of the size of the
  basis set used in our nonequilibrium calculations and does not affect the
  propagation of the LL solution following the quench, but merely limits the
  accuracy with which we can extract correlations from the solutions. In
  practice, we always choose the Cartesian grid density to be sufficiently high
  that the characteristic $\propto k^{-4}$ scaling of the high-momentum tail is
  observed over a broad range of momenta.}
\bibitem{NoteL}
{In practice, we find that the results we obtain for the overlaps from our
  method~\cite{Zill2014} agree with the recently derived analytical expressions
  for these
  quantities~\cite{DeNardis2014a,Brockmann2014a,Brockmann2014b,Brockmann2014c},
  which imply, in particular, that $\langle \{\lambda_j\} | \psi_0 \rangle \sim
  1/\lambda_j^2$ as any $\lambda_j\to\infty$.}
\vspace{0.05\baselineskip}
\end{thebibliography}

\end{document}